\documentclass[prb,
twocolumn,
nopacs,
nofootinbib,
superscriptaddress]{revtex4-2}
\usepackage{bm}
\usepackage{physics}
\usepackage{graphicx}
\usepackage{subcaption} 
\usepackage{booktabs}
\usepackage{amsmath, amssymb}

\usepackage[colorlinks,
    linkcolor=blue,
    anchorcolor=red,
    citecolor=red
]{hyperref}
\usepackage{wasysym}
\usepackage{MnSymbol}
\usepackage{comment}
\usepackage[dvipsnames]{xcolor}
\usepackage[normalem]{ulem}\usepackage[normalem]{ulem}

\usepackage{wasysym}
\usepackage{caption}
\usepackage{ragged2e}

\renewcommand{\v}[1]{\mathbf{#1}}

\begin{document}
\global\long\def\P{\mathcal{P}}%
\global\long\def\u{\upsilon}%

\title{First order
vortex-unbinding transition in 2D topological superconductors}

\author{Vera Mikheeva}
\affiliation{Nanocenter CENN, Ljubljana, Slovenia}
\affiliation{Faculty of Mathematics and Physics, University of Ljubljana, Jadranska 19, 1000 Ljubljana, Slovenia}

\author{Elio J. K\"onig}
\affiliation{Department of Physics, University of Wisconsin-Madison, Madison, Wisconsin, USA}

\begin{abstract}
We study the fate of the Berezinskii--Kosterlitz--Thouless (BKT) transition in a two-dimensional topological superconductor whose vortex cores each host an isolated Majorana zero mode, energetically well separated from the bulk gap and other bound states of the vortex. When the Majorana localization length $\xi_F$ exceeds the vortex core size $\xi$, nearby vortices acquire an additional short-range attractive channel from Majorana hybridization, modifying the vortex-antivortex interaction at short distances and generating a metastable minimum in the two-vortex free energy above the nominal $T_{BKT}$. We construct an effective field theory for the resulting dilute vortex gas, coupling the ordinary BKT phase field to an additional field encoding the Majorana-mediated interaction, 
and show, using renormalization-group and instanton methods, that this hybridization can drive the transition first order, with the metastable phase decaying at an exponentially small rate. We propose an array of Josephson-coupled $p+ip$ superconducting islands as a microscopic platform realizing the required hierarchy of scales. 
Finally, we discuss experimental signatures of this first-order transition, in particular a hysteresis loop whose width is set by the coexistence window.
\end{abstract}

\maketitle

\section{Introduction}
\label{sec: introduction}
Superconductivity in two spatial dimensions is destroyed not by thermal 
fluctuations of
the order parameter amplitude, but by the unbinding of vortex--antivortex pairs at the Berezinskii--Kosterlitz--Thouless (BKT) transition~\cite{berezinskii1971destruction,kosterlitz1973ordering}. This transition is universal: it is controlled entirely by the logarithmic interaction between vortices on length scales larger than
the vortex core size $\xi$ (the coherence length) rather than by any microscopic detail of the pairing mechanism. In a conventional, weakly coupled superconductor, the physics inside the vortex core itself plays no special role for
the transition. A vortex traps a set of subgap Andreev bound states, the Caroli--de~Gennes--Matricon (CdGM) levels~\cite{caroli1964bound}, with a level spacing $\sim \Delta^2/\epsilon_F$ (with $\Delta$ the superconducting gap and $\epsilon_F$ the Fermi energy) 
far smaller than the energy scale associated with the transition temperature $T_{BKT}$:
 the core is thus effectively gapless, behaving as a thermally equilibrated small metallic island rather than a quantum-coherent object with resolvable internal structure. Consequently, 
the vortex-core 
physics
only sets ultraviolet values of coupling constants, but otherwise does not enter the BKT problem.
These arguments also hold for topological superconductors in the weak coupling regime, even if --- as, e.g., for chiral $p+ip$ pairing --- a vortex core hosts a 
Majorana zero mode (MZM)~\cite{alicea2012new, volovik1999fermion, ivanov2001non}. As before, in this limit, the minigap between the Majorana mode and other CdGM states is small compared to the thermal broadening, leading to a featureless metallic core.
\begin{figure}
    \centering
    \begin{subfigure}[b]{0.48\columnwidth}
        \includegraphics[width=\textwidth]{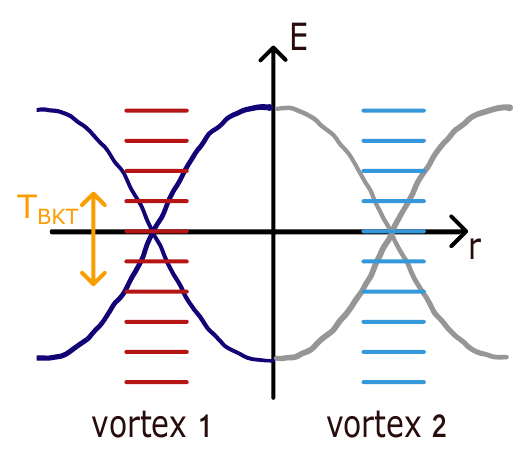}
        \caption{}
        \label{fig: subgap weak}
    \end{subfigure}
    \hfill
    \begin{subfigure}[b]{0.48\columnwidth}
        \includegraphics[width=\textwidth]{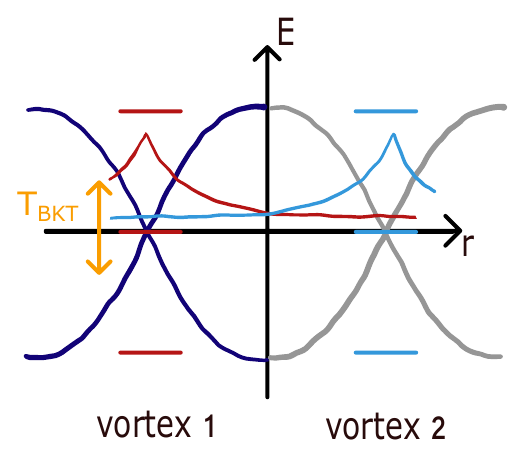}
        \caption{}
        \label{fig: subgap strong}
    \end{subfigure}

    \vspace{0.5em}

    \begin{subfigure}[b]{\columnwidth}
        \includegraphics[width=\textwidth]{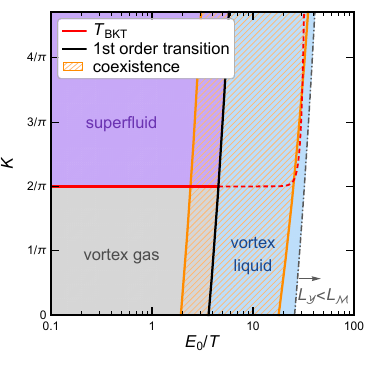}
        \caption{}
        \label{fig: phase diagram inset}
    \end{subfigure}
    \caption{
    \justifying
    Subgap spectrum bound to two vortices (a) for weak-coupling superconductivity and (b) in the limit of strong-coupling topological superconductivity. If the wave function of MZMs decays on a scale longer than the vortex core, the Majorana hybridization leads to an additional energetic gain resulting in short range attraction of vortices.
    (c) Phase diagram in the $(E_0/T, K)$ plane (logarithmic $E_0/T$ axis) for bare vortex fugacity $\mathcal{Y}(0)=10^{-7}$ and $\xi_F/\xi=5$. Shading indicates the global minimum: superfluid (purple) and vortex gas (gray) are separated by the red BKT transition line, while the vortex liquid is separated from either other phase by a first order line (black) and an associated coexistence region (orange hatched). For further details, see Sec.~\ref{sec:PhaseDiagDiscussion}, below.}
    \label{fig: phase diagram}
\end{figure}

In this work, we instead consider topological superconductors in the quantum limit of subgap states.
Unlike the featureless CdGM continuum of a 
vortex of weak-coupling superconductors, we assume that the MZM is an energetically isolated, well-defined mode. 
This means that the
minigap $E_0$ separating it from the first excited core state exceeds the temperature: $E_0>k_B T_{BKT}$ (we set Boltzmann's constant $k_B = 1$ in what follows). 
The MZM wavefunction decays into the bulk over a characteristic length
$\xi_F$ set by the fermionic bulk gap and the 
fermion
velocity, which is
in general distinct from, and can be parametrically larger than, the vortex
core size $\xi$ itself, see Fig.~\ref{fig: phase diagram}\subref{fig: subgap strong}.
Whenever this hierarchy holds, two vortices develop an additional short-range attractive channel well before their cores begin to overlap: their Majorana modes hybridize once the vortex separation reaches $\xi_F$, splitting the zero-energy manifold into two levels at $\pm E_{\rm v\text{-}av}(r)$~\cite{cheng2009splitting} and generating an energy gain that is entirely absent in the classical
vortex problem.

Experimentally, this regime is no longer purely hypothetical. Joining a broader family of unconventional superconductors realized in twisted 2D materials~\cite{cao2018unconventional,park2021tunable,hao2021electric,xia2025superconductivity,guo2025superconductivity,xu2025signatures}, rhombohedral multilayer graphene has emerged as a platform for unconventional superconductivity at very low carrier density~\cite{zhou2021superconductivity,zhou2022isospin,zhang2023enhanced,han2024correlated,nguyen2026coexisting,hua2026multi,kalantre2026fermiology,han2025signatures, dutta2026reconfigurable}  
culminating in reported signatures of chiral, time-reversal-symmetry-breaking pairing 
--- a natural candidate for topological superconductivity with Majorana-carrying vortices. Crucially, 
in this system  
the coherence length and 
the interparticle spacing 
are comparable scales, the opposite of the conventional weak-coupling hierarchy in which vortex-core physics is irrelevant to the BKT transition. Thermal hysteresis, a hallmark commonly associated with first-order rather than continuous transitions, has separately been observed around the superconducting transition in devices of the same materials family~\cite{qin2025stripe}, where it was interpreted as a signature of competing orders.
While the first order physics discussed in this paper need not be the origin of that particular observation, it illustrates that 
first-order-like phenomenology is not exotic to this platform. 
Generally, superconductivity in rhombohedral graphene emerges within an exceptionally rich landscape of correlated phases, including spin- and valley-polarized metals~\cite{zhou2021half,koh2024correlated, geier2025chiral, herasymchuk2026correlated}, anomalous Hall states with both intrinsic and disorder-induced contributions~\cite{nagaosa2010anomalous,ado2015anomalous,ado2016anomalous,mikheeva2026anomalous}, Wigner and metallic Wigner crystals~\cite{han2026evidence,dong2026crystals}, and integer and fractional quantum anomalous Hall phases~\cite{lu2024fractional,lu2025extended,xie2025tunable}. Recent experiments have further reported chiral superconductivity in rhombohedral tetralayer and pentalayer graphene, with superconductivity emerging from spin- and valley-polarized quarter-metal states and exhibiting signatures of spontaneous time-reversal-symmetry breaking~\cite{han2025signatures}. Theoretical work has predicted a spontaneously formed vortex-antivortex lattice hosting Majorana zero modes in the rhombohedral graphene phase diagram~\cite{gaggioli2025spontaneous}, further motivating Majorana hybridization as a nontrivial contribution to vortex energetics.
This phenomenology is not unique to graphene: moiré transition-metal dichalcogenides, most notably twisted MoTe$_2$, have likewise emerged as platforms hosting superconductivity in close proximity to integer and fractional quantum anomalous Hall states~\cite{xu2025signatures}, suggesting that the mechanism explored here may be of broader relevance to flat-band superconductors with intertwined topological and correlated orders.

The above
examples illustrate that a superconducting transition can become first order  when an additional energetic scale or
internal degree of freedom qualitatively modifies the free-energy landscape. This motivates the central question of this paper: can energetically isolated Majorana zero modes, hybridizing between nearby vortex cores,  
alter the character and phase boundary of the BKT transition? 

We answer this question by first analyzing the two-vortex problem, where
Majorana hybridization modifies the vortex-antivortex interaction at short
distances and generates a metastable local minimum in the pair free energy
above the nominal $T_{BKT}$, Fig.~\ref{fig: vortex_dipole_F}. We then extend
this to the many-vortex problem, constructing an effective field theory that
couples the ordinary BKT phase field to an additional field encoding the
Majorana-mediated interaction. 

From this we derive an effective phase diagram which features {a dense vortex liquid stabilized by the Majorana hybridization.} A renormalization-group and instanton analysis of this theory shows that, in the appropriate parameter regime, the transition does indeed become first order (Fig.~\ref{fig: phase diagram inset}).
This is structurally analogous to false-vacuum decay via bubble nucleation, a phenomenology that has recently been probed experimentally in a different setting --- a neutral, ferromagnetic atomic superfluid~\cite{zenesini2024false} --- using measurement protocols adapted from macroscopic quantum tunneling experiments in Josephson junctions~\cite{devoret1985measurements,martinis1987experimental}. While the underlying physics and system there are distinct from the superconducting vortex gas considered here, 
that work demonstrates that the decay of a metastable phase is experimentally accessible with this class of technique once the nucleation barrier becomes sufficiently small, i.e.\ close to the limit of metastability.  Note that BKT physics can be expected even in charged 2D superconductors, as the system size in realistic thin-film or 2D samples is typically much shorter than the Pearl length~\cite{pearl1964current}.

First-order transitions within otherwise BKT-type problems have also been noted in purely classical settings. Monte Carlo studies of the two-dimensional generalized XY model~\cite{mol2006monte,dasilva2024firstorder} have shown that modifying the nearest-neighbor spin-spin interaction via a generalization parameter $q$ drives the transition from continuous (BKT) to first order beyond a critical $q_c$, effectively renormalizing the vortex-core energetics. Separately, a Monte Carlo study of the 2D Coulomb gas at fixed density~\cite{orkoulas1996phase} found a first-order vapor--liquid coexistence line terminating the BKT line, a distinct, density-driven instability. In both cases the relevant parameter is externally imposed rather than derived from a microscopic mechanism. 

We conclude this introduction with a brief digression to $(1+1)$-dimensional wires, for which the superconducting quantum phase transition is usually also BKT-like. However, for topological superconductors, Majorana bound states centered at the space-time position of a phase slip are known to alter the universality class \cite{shi2011boson, roy2020quantum,   bollmann2026phases}  of the quantum phase transition or even stabilize \cite{rampp2022topologically} the superconducting phase.

The remainder of this paper is organized as follows. In Sec.~\ref{sec: two vortex problem} we introduce the two-vortex problem, deriving the modified vortex-pair free energy from Majorana hybridization and identifying the hierarchy of length and energy scales required for our analysis.
In Sec.~\ref{sec:MicroscopicModels} we discuss a concrete microscopic platform realizing this hierarchy: an array of Josephson-coupled $p+ip$ superconducting islands.
In Sec.~\ref{sec: EFT} we construct the effective field theory for the resulting dilute vortex gas. In Sec.~\ref{sec: RG} we analyze its renormalization-group flow and determine the resulting phase diagram, Fig.~\ref{fig: phase diagram inset}. In Sec.~\ref{sec: instantons} we compute the decay rate of the metastable phase using instanton methods and finally conclude with a discussion and outlook. Technical details are relegated to the appendices.

\section{Two vortex problem}
\label{sec: two vortex problem}
As a physical preamble for this problem, we begin by considering the simplest nontrivial vortex configuration ---
a single vortex-antivortex pair. This problem captures the elementary object responsible for the BKT transition
while simultaneously allowing us to incorporate the interaction between Majorana bound states localized in the vortex cores. It is worth reminding the reader that the BKT transition occurs in the extremely dilute limit of vortices (technically speaking, the critical renormalization group end point is at zero fugacity). Therefore, it is well motivated to first study 
the effective free energy of a vortex pair  
as the starting point for the subsequent thermodynamic analysis.

Throughout this work we consider a strongly coupled
topological superconductor satisfying the hierarchy: 
\begin{align}
    E_0 > T_{\text{BKT}},
    \label{eq: E/T condition}
\end{align}
 where 
 $E_0$ denotes the Caroli–de~Gennes–Matricon (CdGM) minigap separating the Majorana zero mode from the first excited vortex-core states or the fermionic continuum, see Figs.~\ref{fig: phase diagram}\subref{fig: subgap weak} and \subref{fig: subgap strong}. Under these conditions a single vortex hosts an isolated Majorana zero mode, while higher CdGM states remain thermally inaccessible. Furthermore, we assume that the vortex core size, $\xi$, is much smaller than the characteristic localization length of the Majorana bound state, $\xi_F$: 
 \begin{align}
\xi_F \gg \xi,
     \label{eq: xi condition}
 \end{align}
ensuring that the Majorana wave functions overlap while the vortex cores remain well separated.  Leaving details to Sec.~\ref{sec:MicroscopicModels}  below, such a scenario requires a bosonic superconducting mechanism which sets $\xi$, while fermions are quasi-gapless with large $\xi_F$ (e.g. close to a topological phase transition).

\begin{figure}[t]
    \centering
    \includegraphics[width=1\linewidth]{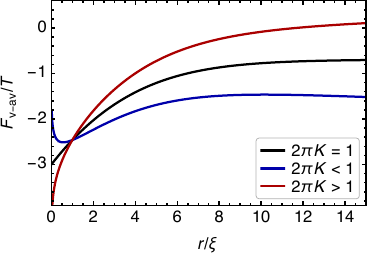}
    \caption{ 
        \justifying Vortex--antivortex free energy $F_{\rm v\text{-}av}/T$, Eq.~\eqref{eq: F_v/av}, as a function of vortex separation $r/\xi$, for $2\pi K = 1$ (black), $2\pi K < 1$ (blue), and $2\pi K > 1$ (red), at fixed $E_0/T = 3$ and $\xi_F/\xi = 5$. All curves cross at $r=\xi$, where the logarithmic term vanishes. At large separations all curves are dominated by the logarithmic stiffness/entropy competition. For $2\pi K=1$ the free energy approaches $-T \ln 2$, the entropy of the fermionic state formed by the two Majorana modes. At short distances, the Majorana correction lowers the free energy. As a result, for $2\pi K < 1$ the free energy displays two minima even above the nominal BKT unbinding temperature: a metastable one at short distances (bound pair) and one at large separations (unbound vortices).}

    \label{fig: vortex_dipole_F}
\end{figure}

Consequently, we model the Majorana hybridization energy of a vortex–antivortex pair separated by a distance $r$ by the following form:
\begin{align}
E_{\text{v-av}}(r)=E_0 e^{-r/\xi_F}.
    \label{eq: E_v-av}
\end{align}
This hybridization splits the zero-energy state into two fermionic levels with energies $\pm E_{\text{v-av}}(r)$. The many-body Hamiltonian of the two Majorana modes $\gamma_{1,2}$ centered at the two cores is thus $H = i \gamma_1 \gamma_2 E_{\rm v-av}(r)$ (throughout the manuscript Majorana operators obey $\{\gamma_i, \gamma_j \} = 2 \delta_{ij}$). This gives rise to the fermionic partition function
\begin{align}
  Z_{\mathrm{F}}(r) &= 2 \cosh \left( \frac{E_{\text{v-av}}(r) }{T} \right),
  \label{eq: Zferm}
\end{align}
which enters the statistical weight of a single vortex--antivortex pair: 
\begin{align}
    Z_{\text{v-av}}
   &= e^{-2E_{\rm core}/T } \int \frac{d^2 R\,  d^2 r}{\xi^4}
       Z_{\mathrm{F}}(r)\,
       e^{-2\pi K\ln\left(\frac{r}{\xi}\right)} \notag\\
    & = e^{-2E_{\rm core}/T }\frac{ \mathcal A}{\xi^3} \int dr\, e^{- F_{\rm v-av}(r)/T}
\end{align}
where $e^{-E_{\rm core}/T}$ is the Boltzmann weight associated with the core energy of a single vortex, $K$ is the dimensionless stiffness obtained by dividing the superfluid stiffness $J$ by temperature, $K = J/T$, and $\mathcal A$ is the area of the system.
In the second line we
introduced the free energy of a single vortex--antivortex pair as a function of their separation:
\begin{align}
F_{\text{v-av}}(r)=T(2\pi K-1)\bigl(\ln r - \ln \xi\bigr)&
           \notag \\- T\ln \left[2 \cosh \left( \frac{E_0}{T} e^{-r/\xi_F} \right) \right].
    \label{eq: F_v/av}
\end{align}
The first line encodes the basic phenomenology of the BKT transition: the term proportional to stiffness is attractive, while the term $- T \ln(r/\xi)$ stems from the measure and is an entropic, effectively repulsive contribution. By consequence, the sign of the logarithmic potential changes when energetic and entropic contributions are comparable, $2\pi K = 1$, corresponding to the vortex unbinding transition temperature, which in this simplified procedure is overestimated. While at large separations the logarithmic interaction always dominates, the Majorana contribution (second line) lowers the free energy at short
distances and may generate a local minimum even above the nominal vortex
unbinding transition, as illustrated in Fig.~\ref{fig: vortex_dipole_F}. 

The physical origin of the emergent minimum is the fermionic energy gain associated with the hybridization of the Majorana modes as the vortices approach each other.  The above analysis relies on the hierarchy $\xi_F \gg \xi$, which ensures that Majorana hybridization takes place while the vortices remain well-defined objects. {We emphasize that this two-vortex calculation is an illustration of the mechanism rather than a quantitative criterion. What carries over to the many-vortex problem is the short-range attraction of strength $E_0$ and range $\xi_F$, which is independent of vorticity and, acting between all vortices, drives the first-order transition derived below}. In the following section, we discuss physical systems in which this hierarchy can be realized before returning to the many-vortex problem describing the statistical physics of vortices at the BKT transition later in the manuscript.

\section{Microscopic model: Josephson-coupled $p+ip$ island array}
\label{sec:MicroscopicModels}

One platform realizing  
the hierarchy of scales of interest is provided by Josephson junction
systems. In this section, we introduce the microscopic model of the
junction and derive the effective description of the Majorana degrees of
freedom. Tight-binding lattice models built from Majorana modes have
previously been used to describe hybridization between vortex-bound
Majorana modes in $p$-wave superconductors and topological-insulator
heterostructures~\cite{Biswas2013,LiuFranz2015}, and, in a closely
related but distinct setting, to describe the coupling of Majorana modes
confined to mesoscopic superconducting islands~\cite{OregVonOppen2020,
BravyiTerhalLeemhuis2010}. Here we adapt this approach to a Josephson
junction array, in which the relevant vortex is a flux  
threading
the array rather than an order-parameter vortex in the bulk of a $p$-wave
superconductor; the vortex core size in this setting is set by the array
lattice spacing itself, analogous to the fluxon core size in more
conventional Josephson junction arrays~\cite{Newrock2000}.

\begin{figure}[t]
    \centering
    \includegraphics[width=.8\linewidth]{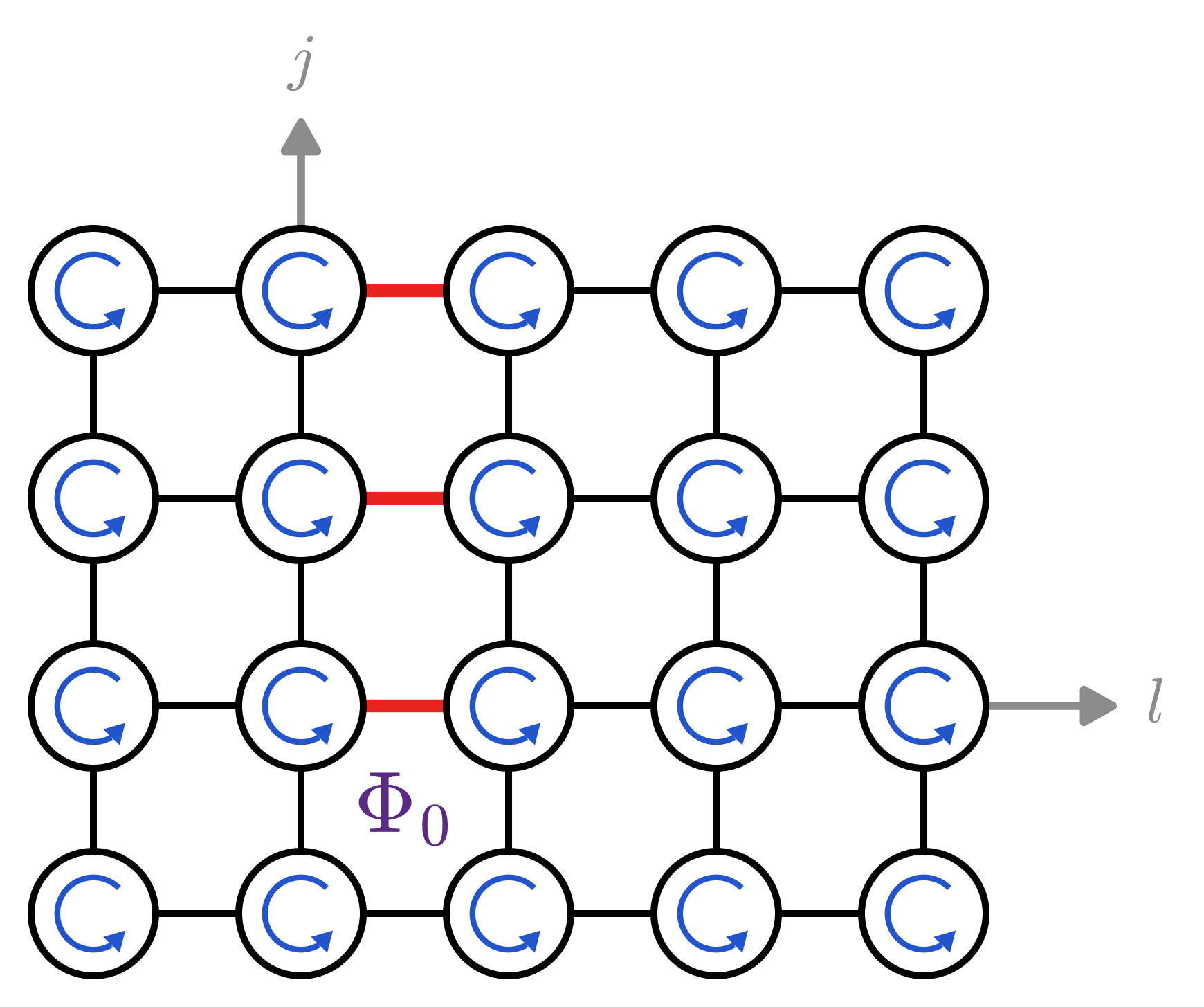}
    \caption{
    \justifying  Schematic Josephson-junction array hosting a square
    lattice of Majorana modes, pierced by a single flux quantum $\Phi_0$.
    Axes $l$ (right) and $j$ (up) follow the indexing of
    Eq.~\eqref{eq: JJ final Hamiltonian with flux}. The vortex core sits
    in the plaquette with corners $(l,j)=(0,-1),(0,0),(1,-1),(1,0)$. The
    branch cut (red, $-1$) runs from it along $j\geq0$; bonds crossing it
    acquire the sign $s(l,j)=-1$ of Eq.~\eqref{eq: branch cut}, all other
    bonds have $s=+1$.}
    \label{fig: JJ}
\end{figure}

We consider an array of mesoscopic superconducting islands, each
represented by a disk of radius $R$ made of a two-dimensional chiral
$p+i p$ superconductor, Fig.~\ref{fig: JJ}. 
The islands are coupled via point-like tunneling
contacts. For concreteness, we assume that each disk is connected to its
neighbors through four contacts located at its boundary and oriented
along mutually orthogonal directions.

We remind the reader that the chiral boundary mode of a $p+ip$-wave superconductor is subject to antiperiodic boundary conditions~\cite{FendleyNayak} leading to finite size quantized momentum $p_m = (m + 1/2)/R$, $m \in \mathbb Z$. For theoretical convenience, we assume the dispersion $\epsilon(p_m)$ such that the modes $m = -1,0$ have negligible $\vert \epsilon(p_m )\vert$ and are energetically separated from all other boundary modes. By particle-hole symmetry of the BdG spectrum, these two channels are particle-hole partners of the same mode, not independent degrees of freedom — one being near zero energy guarantees the other is too. This single near-zero-energy channel is described by one complex fermionic operator $c_0$, which -- like any complex fermion -- decomposes into two Hermitian (Majorana) components, $c_0=(\gamma_0+i\gamma_0')/2$.
Each island thus effectively hosts two Majorana zero modes bound to its boundary, arising
from a band structure of chiral edge states that is detailed in
Appendix~\ref{ap: JJ}. Integrating out the bulk of each island reduces
the problem to an effective tight-binding model for these boundary
Majorana modes, $\gamma_0$ and $\gamma_0'$, living on the sites of the
resulting square lattice, shown schematically in
Fig.~\ref{fig: JJ}, where $l$ and $j$ are integer indices labeling the column and row of the
islands, respectively, 
with lattice spacing set to
$a=1$. The full microscopic derivation of this
reduction — starting from the tunneling Hamiltonian between the islands
and proceeding to the low-energy Majorana theory — is given in
Appendix~\ref{ap: JJ}; here we only state the result needed for the rest
of the discussion.

Piercing the array with a flux $\Phi_0$ requires choosing a branch cut,
or Dirac string, along which the induced Peierls phase is concentrated;
away from the string, bonds are left unmodified. We choose the string to
run along the bonds connecting column $l=0$ to column $l=1$ for $j\geq0$, so that the phase winds by $2\pi$ around
any plaquette encircling the vortex core, located in the plaquette
between columns $l=0,1$ and rows $j=-1,0$. This choice is
encoded in the sign function $s(l,j)$ of Eq.~\eqref{eq: branch cut}:
bonds crossing the string acquire $s=-1$, while all other bonds have
$s=+1$. Because $s(l,j)$ only ever enters as a relative sign between the
two gauge sectors, any other choice of string geometry terminating on the
same vortex is related to this one by a local redefinition of the
$\gamma_0'$ operators along the alternative path, and leaves the spectrum
unchanged. The effective Hamiltonian for the coupled Majorana
modes then takes the form
\begin{align}
H_{\Phi_0} = &-i\tilde{t}\sum_{l,j}\Big[
\gamma_0'({l,j})\gamma_0'({l,j+1})
- \gamma_0({l,j})\gamma_0({l,j+1}) \notag \\
&- \gamma_0'({l,j})\gamma_0({l,j+1})
+ \gamma_0({l,j})\gamma_0'({l,j+1})
\Big] \notag \\
&- 2i\tilde{t}\sum_{l,j} s(l,j)\,
\gamma_0({l,j})\gamma_0'({l+1,j}) \notag \\
&+ 2iM\tilde t \sum_{l,j}
\gamma_0'({l,j})\gamma_0({l,j}),
\label{eq: JJ final Hamiltonian with flux}
\end{align}
with
\begin{equation}
s(l,j) =
\begin{cases}
+1, & l\neq 0,\\
-1, & l=0,\ j\geq 0,\\
+1, & l=0,\ j<0.
\end{cases}
\label{eq: branch cut}
\end{equation}
The sign change of $s(l,j)$ along the line $l=0$, $j\geq0$ represents the
branch cut associated with the inserted flux. In the absence of a vortex,
there is no branch cut and $s(l,j)\equiv+1$ for all $(l,j)$, so that the
Hamiltonian is translationally invariant.

The first two lines of
Eq.~\eqref{eq: JJ final Hamiltonian with flux} describe the hopping of
Majorana modes between neighboring islands along the $j$-direction, while
the third line describes hopping along the $l$-direction. The parameter $\tilde t$ is the effective
tunneling amplitude between neighboring Majorana modes, inherited from
the microscopic contact transparency. The last term is an on-site
hybridization between the two Majorana modes on the same island due to the finite extent of the islands, with $M$
controlling the topological phase of the array.

We use open boundary conditions in both spatial directions. The vortex is
placed in the bulk of the finite lattice, and the corresponding branch
cut terminates at the boundary. We then use this Hamiltonian to
numerically search for Majorana zero modes, restricting $M$ to the window
$-2<M<0$, which follows from requiring a nontrivial Chern number for the
corresponding lattice model in the absence of the flux (see
Appendix~\ref{ap: JJ}).

Fig.~\ref{fig: MZM decay length} 
\begin{figure}[t]
    \centering
    \begin{subfigure}[b]{0.49\linewidth}
        \centering
        \includegraphics[width=\linewidth]{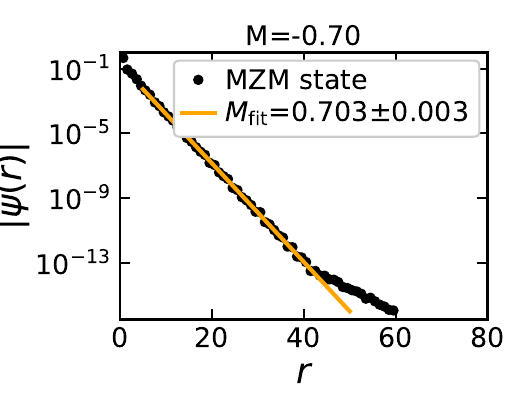}
        \caption{}
        \label{fig: MZM density example}
    \end{subfigure}
    \hfill
    \begin{subfigure}[b]{0.49\linewidth}
        \centering
        \includegraphics[width=\linewidth]{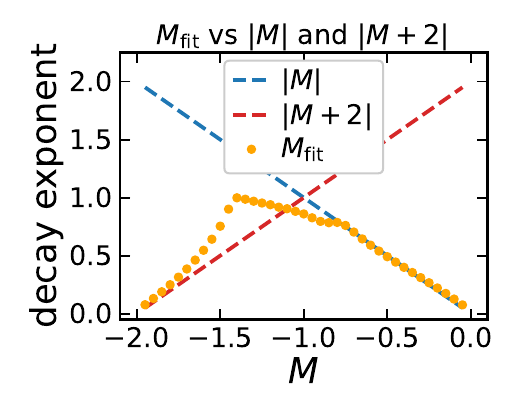}
        \caption{}
        \label{fig: Mfit}
    \end{subfigure}
    \caption{
    \justifying
(a) Radial profile $|\psi(r)|$ of the Majorana zero-mode wavefunction as
    a function of the distance $r$ from the vortex core, averaged over
    shells in the half-plane away from the branch cut and shown on a
    semi-logarithmic scale, for a representative point in the topological
    window, $M=-0.7$ (black points), together with an exponential fit
    for $5\leq r\leq40$ (orange line). For $r\gtrsim40$ the amplitudes along the diagonals, which decay
    faster on the lattice, fall below the numerical accuracy and are
    excluded from the shell average, the profile is then dominated by
    the slower decay along the axes, which changes the slope.
(b) Decay rate (inverse decay length) $M_{\rm fit}$ of the Majorana
    wavefunction as a function of $M$ across the window $-2<M<0$,
    extracted from a direct exponential fit to the numerical
    wavefunction at each $M$ (orange points). Near
    $M\to0^-$ the decay follows $\sim e^{-|M|r}$ (blue dashed line),
    set by the bulk gap closing at the $X/Y$ points; near $M\to-2^+$
    it follows $\sim e^{-|M+2|r}$ (red dashed line), set by the gap
    closing at \(\Gamma\).}
    \label{fig: MZM decay length}
\end{figure}
shows the spatial decay of the
Majorana wavefunction and its decay rate as a function of $M$ across this window. Near
$M\to0^-$, the decay follows $\sim e^{-|M|r}$, as expected since the bulk
gap closes at the $X/Y$ points of the Brillouin zone in this limit. Near
the opposite end of the window, $M\to-2^+$, the decay instead follows
$\sim e^{-|M+2|r}$: the two ends of the topological window correspond to
distinct bulk gap closings, so the localization length is controlled by
$M$ near one and by $M+2$ near the other. The orange points in Fig.~\ref{fig: MZM decay length}\subref{fig: Mfit}
show the exponent $M_{\rm fit}$ extracted from directly fitting the
numerical Majorana wavefunction to $\sim e^{-M_{\rm fit}r}$. Away from the two limits $M_{\rm fit}$ is not simply proportional to the
bulk gap $E_0$: there the decay rate is given by the imaginary momentum at
which the bulk energy of
Eq.~\eqref{eq: JJ final Hamiltonian with flux} vanishes, rather than by
the gap minimum, and $M_{\rm fit}$ averages the anisotropic decay over
directions.

This result has a direct physical consequence for the platform: the
numerically extracted decay length of the Majorana wavefunction,
$\xi_F=1/M_{\rm fit}$, diverges as $1/|M|$ for $M\to0^-$ and as $1/|M+2|$ for $M\to-2^+$, while the lattice spacing $a$ plays the role of the vortex core size $\xi$.
Since $M$ can be tuned continuously,
one can access the regime $\xi_F\gg\xi$ simply by tuning $M$ close
to either end of the topological window, without requiring any additional
fine-tuning of the microscopic parameters of the junction.

The same numerics also yield the spectrum of vortex-bound states, shown in Fig.~\ref{fig: spectrum ap} of Appendix~\ref{ap: JJ}. Apart
from the MZM, no other state below the bulk gap is localized at the
vortex: since the vortex core has the size of a single plaquette, the
Caroli--de Gennes--Matricon levels are pushed into the continuum. The
separation between the MZM and the next vortex-bound level is therefore
the full bulk gap,
$E_0=4 \tilde t\min(|M|,|M+2|)$, e.g.\ $E_0=2.8\,\tilde t$ for
$M=-0.7$.
This energy has to be compared with $T_{\rm BKT}\sim\,E_J$, set by
the Josephson coupling $E_J$ between neighboring
islands~\cite{hasenbusch2005two}. While $E_J$ is second order in the
tunneling amplitude of the contacts ~\cite{altland2010condensed}, 
$\tilde t$ is linear in it (Appendix~\ref{ap: JJ}), so that
$E_0/T_{\rm BKT}\sim\tilde t/E_J\gg1$ for weakly transparent contacts.
    
\section{Effective field theory}
\label{sec: EFT}
Building on the two-vortex analysis above, we now consider a dilute
gas of vortices and antivortices, 
and present the full analysis and calculations for the vortex unbinding 
transition in the presence of Majorana-mediated vortex interactions. As a first step, in this section we derive an effective field theory.
\subsection{Assumptions}
We control our calculations
throughout by concentrating on the dilute limit
\begin{equation}
n\xi_F^2\ll1,
\end{equation}
where $n$ is the vortex concentration and $\xi_F$ is the Majorana localization length introduced in Sec.~\ref{sec: two vortex problem}. Since $\xi_F\gg\xi$ in the hierarchy of Eq.~\eqref{eq: xi condition}, this is a stronger condition than the  
requirement that vortex cores do not overlap, $n\xi^2\ll1$: it further ensures that a typical vortex has, on average, no more than one partner within the range of the Majorana-mediated hybridization interaction introduced below. 
This is partly motivated by
the usual BKT treatment, where the leading dilute-gas description is formulated in terms of vortex–antivortex pairs, while larger neutral clusters can be generated but do not need to be treated as independent elementary objects at leading order (recall that the BKT critical end point occurs at zero fugacity).

Here we similarly retain the two-vortex sector, with the vortex–antivortex Majorana hybridization described in Sec.~\ref{sec: two vortex problem}, and neglect higher-order multi-vortex clusters. 
Disregarding configurations with more than two vortices within range $\xi_F$ allows us to simplify the fermionic partition sum in the presence of $N$ vortices and $N$ antivortices at fixed positions $\{\v x_i\}_{i = 1}^{2N}$. To be concrete, we denote the number of vortex pairs that are mutually within range $\xi_F$ of each other by $N_{\rm pair}$ and number the vortex positions such that $\v x_1$ is close to $\v x_2$, $\v x_3$ close to $\v x_4$ and so forth up to $\v x_{2 N_{\rm pair}}$. Then,
\begin{align}
    Z_F(\{\v x_i\}) &\simeq 2^{N-N_{\rm pair}} \prod_{i = 1}^{N_{\rm pair}} Z_F(\vert \v x_{2i-1}- \v x_{2i} \vert) \notag \\
    &\simeq 2^{N} \prod_{\v x \neq \v x'} \sqrt{\frac{Z_F(\vert \v x - \v x' \vert )}{2}}. \label{eq:ApproxZF}
    \end{align}
In the last line we extended the product to run over positions $\v x, \v x'$ of all $2N$ vortices and we use that most factors are exponentially close to unity ${Z_F(r)}/{2} \stackrel{r   \gg \xi_F}{\simeq }1 + \tfrac12 (E_0/T)^2 e^{-2r/\xi_F}$ and thus do not contribute. The square root compensates double-counting. In particular, if all vortices are farther apart than $\xi_F$, $Z_F(\{\v x_i\}) \simeq 2^N$ reproduces the $2^N$-fold degeneracy of the $2N$ Majorana modes.

\subsection{The model}
\label{sec: the model}
Building on the discussion of the previous section, we describe the dilute gas of vortices and antivortices as a lattice gas of unit
charges $q_{\mathbf r}=0,\pm1$ on cells of area $\xi^2$, labeled by
their positions $\mathbf r$; the sum over particle number and the integral
over positions of the continuum problem are thereby replaced by a sum
over local occupation numbers $\{q_{\mathbf r}\}$, as a concrete UV regularization scheme of the field theory.
The resulting grand-canonical
partition sum is
\begin{align}
\mathcal{Z}_G&=\sum_{\{q_{\mathbf r}=0,\pm1\}}\exp\Big[
\pi K\sum_{\mathbf r\neq\mathbf r'}q_{\mathbf r}q_{\mathbf r'}\,
   \ln\frac{|\mathbf r-\mathbf r'|}{\xi}\notag\\
&+\frac1{2T}\sum_{\mathbf r\neq\mathbf r'}|q_{\mathbf r}|\,|q_{\mathbf r'}|\,
   V^{\rm ferm}(\mathbf r- \mathbf r')
+\frac{\mu}{T}\sum_{\mathbf r}|q_{\mathbf r}|\Big],
\label{eq: ZG lattice gas}
\end{align}
where the three terms have distinct physical origins. The first is the
familiar logarithmic Coulomb-gas interaction between vortices, with
$K$ the usual dimensionless superfluid stiffness: a single
vortex--antivortex pair costs free energy $2\pi K T\ln(r/\xi)$, as in
Eq.~\eqref{eq: F_v/av}. The second term is new to the present problem:
it is the Majorana-induced hybridization energy between
vortex cores derived in Sec.~\ref{sec: two vortex problem}, written as
an effective pairwise potential, 
\begin{equation}
   V^{\rm ferm}(\mathbf r) =T\ln\cosh\!\left(\frac{E_0}{T}e^{-|\mathbf r|/\xi_F}\right)
=T\ln [Z_{\rm F}(|\mathbf r|)/2],
\label{eq: Vferm def}
\end{equation}
with $Z_{\rm F}(r)$ the fermionic partition function of
Eq.~\eqref{eq: Zferm}; it is this term that distinguishes the present
vortex gas from the standard BKT problem. While the long-range interaction does depend on the sign of the vortex charge (i.e., vorticity), we highlight that the short-range interaction is vorticity-independent. The third term is the
fugacity contribution controlling the density of vortices, with $\mu$
the vortex chemical potential. It contains two contributions. First, as usual, each vortex has a Boltzmann weight $e^{-E_{\rm core}/T}$ associated with the core energy. Second, a system of $N$ vortices and $N$ antivortices contains an overall factor $2^{N}$, see Eq.~\eqref{eq:ApproxZF}. It thus couples to the total density such that the chemical potential obtains a $T$ dependent correction of entropic origin, $\mu = - E_{\rm core} + T \ln(\sqrt{2})$. {Here, $E_{\rm core}$ is understood to include all constant single-vortex contributions, in particular the self-interaction terms generated by the decoupling of the interactions}.

To proceed analytically (see Appendix~\ref{ap: Zeffective} for details), we decouple both interaction terms with two
independent Hubbard--Stratonovich fields. First, the ``disorder field'' $\theta(\mathbf x)$, dual to
the phase field, decouples the logarithmic vortex--vortex interaction. Second, the fluctuating chemical potential $\Psi(\mathbf x)$, which couples to the vortex density analogously to $\mu/T$, decouples the Majorana-mediated interaction. 
Next, we sum over the vortex charges
$\{q_{\mathbf r}\}$, and finally take the continuum limit, replacing the
discrete
lattice positions $\mathbf r$ of the vortex cores by the
 continuum coordinate $\mathbf x$  and the
lattice sum by an
 integral weighted by the inverse cell area,
\begin{equation}
\sum_{\mathbf r}\;\to\;\frac{1}{\xi^2}\int d^2x.
\end{equation}
The result is a generalized sine--Gordon-type action
coupling the disorder field $\theta$ and the fluctuating chemical potential $\Psi$,
\begin{subequations}
\begin{equation}
\mathcal{Z}_G\equiv\int\mathcal{D}\theta\,\mathcal{D}\Psi\,e^{-S[\theta,\Psi]},
\end{equation}
where $S[\theta,\Psi] =S_{\rm C}[\theta] + S_{\rm SR}[\Psi]+ S_{\rm int}[\theta,\Psi] $ and
\begin{align}
S_{\rm C}[\theta]&=\int d^2x\frac{1}{8\pi^2K}(\vec\nabla\theta(\mathbf{x}))^2,\\
S_{\rm SR}[\Psi]&=\frac{T}{2}\int{d^2x} d^2y\,\Psi(\mathbf{x})\,[V^{\rm ferm}]^{-1}(\mathbf{x}-\mathbf{y})\,\Psi(\mathbf{y}),\\
S_{\rm int}[\theta, \Psi]&=-\frac{1}{\xi^2} {\int d^2x} \ln\Big(1+\mathcal{Y}e^{\Psi(\mathbf{x})}\cos(\theta(\mathbf{x}))\Big), \label{eq:IntTerm}
\end{align}
\label{eq: sine-Gordon action}
\end{subequations}
where 
$\mathcal{Y}=2e^{\mu/T}$ is the dimensionless 
fugacity, and
$[V^{\rm ferm}]^{-1}$ is the operator inverse of $V^{\rm ferm}$,
defined by
\begin{equation}
\int d^2z\;V^{\rm ferm}(\mathbf{x}-\mathbf{z})\,[V^{\rm ferm}]^{-1}(\mathbf{z}-\mathbf{y})
=\delta^{(2)}(\mathbf{x}-\mathbf{y}).
\label{eq: Vinv def}
\end{equation}
In Fourier space, $V^{\rm ferm}(\mathbf q)=\int d^2r\,
e^{-i\mathbf q\cdot\mathbf r}\,V^{\rm ferm}(\mathbf r)$ is peaked at
$\mathbf q=0$ with width $|\mathbf q|\sim1/\xi_F$ set by
the same Majorana localization length that controls the real-space decay
of $V^{\rm ferm}(\mathbf x)$.
In the limit of small $\mathcal Y$, Eq.~\eqref{eq: sine-Gordon action} has the same overall
structure as the ordinary sine-Gordon (Coulomb-gas) action controlling
the standard BKT transition, with two differences: the fugacity is not
the bare vortex-core weight and the cosine term is dressed
by $e^{\Psi(\mathbf{x})}$, coupling it to the Hubbard–Stratonovich field  $\Psi$
sourced by $V^{\rm ferm}$. The logarithm entering Eq.~\eqref{eq:IntTerm} is a consequence of the lattice regularization, Eq.~\eqref{eq: appendix discrete effective} of
Appendix~\ref{ap: Zeffective}, and is unambiguous whenever
$1+\mathcal{Y}e^{\Psi(\mathbf{x})}\cos\theta(\mathbf{x})>0$ pointwise.
Where this sign formally flips, Eq.~\eqref{eq: sine-Gordon action} should
be read as the continuum limit of the well-defined lattice product
rather than the literal exponential of Eq.~\eqref{eq:IntTerm} (these
technical subtleties are not of further relevance).

Two dimensionless parameters control the theory: the stiffness $K$ and the ratio $E_0/T$. It is convenient to organize the results below as a $(K,\,E_0/T)$ phase diagram. $K$ alone controls the ordinary BKT unbinding transition. $E_0/T$ measures the strength of the Majorana-mediated attraction relative to thermal fluctuations --- it vanishes in the conventional (non-topological) vortex gas, where Eq.~\eqref{eq: sine-Gordon action} reduces to the standard sine-Gordon action. Increasing $E_0/T$ strengthens the Majorana-mediated interaction and can turn the ordinary continuous BKT transition into a first-order one.

\subsection{Qualitative discussion of phases}
\label{sec:QualDiscPhases}

\begin{figure}
    \centering
    \includegraphics[width=\linewidth]{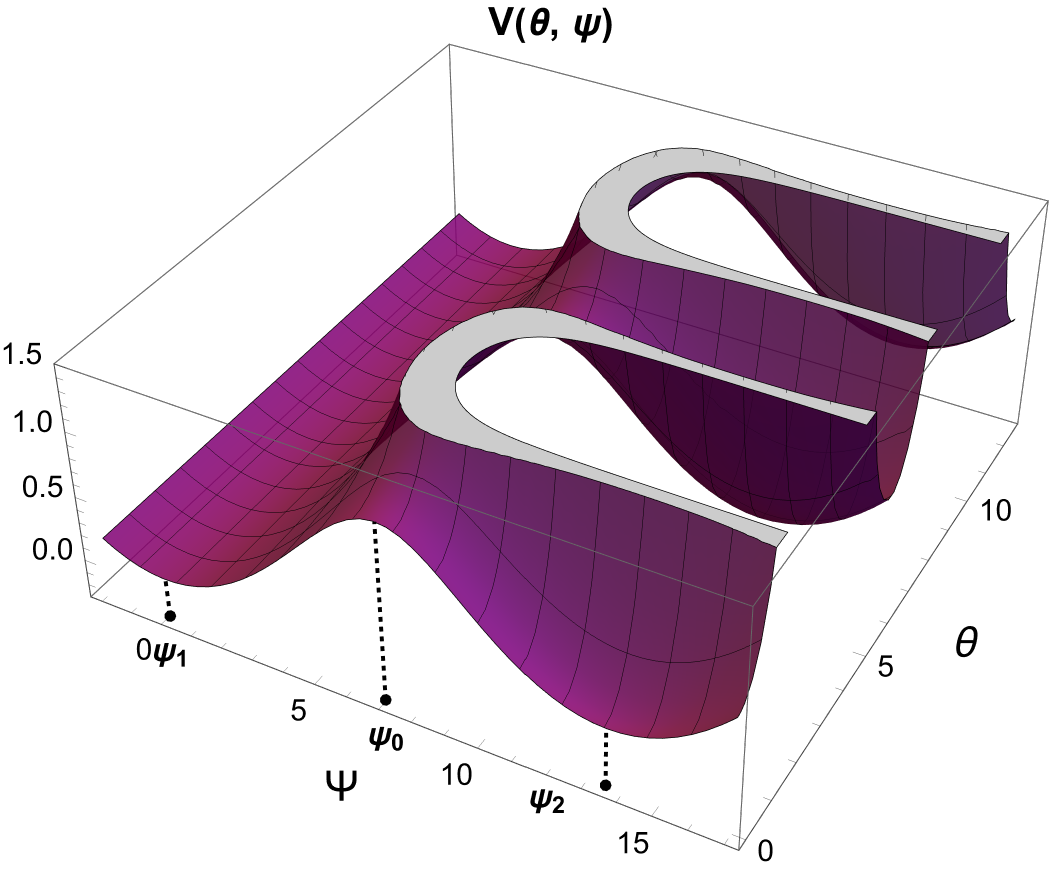}
    \caption{\justifying 
    Potential landscape $V(\theta,\Psi)$, Eq.~\eqref{eq:VthetaPsi}, 
    for $b = 0.037$ and fugacity $\mathcal Y=10^{-3}$. The two minima $\Psi_1$ and $\Psi_2$ are
    nearly degenerate and are separated by the barrier at $\Psi_0$.
    The landscape illustrates the potential structure underlying the
    thin-wall bounce construction below.}
    \label{fig: potential landscape}
\end{figure}

Before turning to the renormalization-group and instanton analyses of the following sections, 
we lay out the main phases expected based on Eq.~\eqref{eq: sine-Gordon action}. For homogeneous fields, the action reduces to $S[\theta, \Psi] = V(\theta, \Psi) \mathcal A/\xi^2$ with the potential 
\begin{equation}
V(\theta,\Psi)= b
\Psi^2-\ln(1+\mathcal Y  e^\Psi {\cos(\theta)}).
\label{eq:VthetaPsi}
\end{equation}
Here, a quadratic (mass-like) term in $\Psi$ with prefactor $b = \frac{T \xi^2}{4\pi E_0 \xi_F^2} \ll 1$ competes 
against the logarithmic term generated by the sum over vortex occupations.
As illustrated in Fig.~\ref{fig: potential landscape}, this potential at $\theta = 0$ develops two generically non-degenerate local minima $\Psi_1<\Psi_2$, 
separated by a barrier at $\Psi_0 \in [\Psi_1,\Psi_2]$ if 
\begin{equation}
b \in [b_<, b_>], \qquad b_< \simeq \frac{e \mathcal Y}{2}, \qquad b_> \simeq \frac{1}{2 \ln(1/\mathcal Y)}. \label{eq:WindowFor2Min}
\end{equation}
Here, the $\simeq$ sign denotes the asymptotic behavior for small $\mathcal Y$, for details see App.~\ref{app:TwoMin}. These conditions imply $(\xi/\xi_F)^2 \ln(1/\mathcal Y)\lesssim E_0/T \lesssim (\xi/\xi_F)^2 /\mathcal Y $.

The two minima correspond to two different phases of vortices in the absence of long range interactions (i.e. at $K = 0$). The short range interactions can stabilize the phase described by the minimum $\Psi_2$, which corresponds to a phase with relatively large vortex density; we call this phase the \emph{vortex liquid}. The phase near $\Psi_1$ in contrast represents the \textit{vortex gas} with relatively low density. The transition between these two phases is first order --- as usual for liquid-gas transitions. The position of this first order transition (i.e. equal minima) occurs at  
\begin{align}
    b_{\text{1st}} &\simeq \frac{1/4}{\ln(1/\mathcal Y)}. \label{eq:FirstOrder}
\end{align}

We emphasize that this is a mean-field result, obtained by comparing the values of the potential at its two minima. The minimum $\Psi_2$, i.e. the vortex liquid, corresponds to a vortex density beyond the dilute limit in which the pairwise form, Eq.~\eqref{eq:ApproxZF}, is justified. However, the loss of stability of the dilute phase at $b_{\text{1st}}$, Eq.~\eqref{eq:FirstOrder}, can be understood from the vortex gas side, i.e. within the regime of validity. Thus, the liquid--gas is  
captured at this level of approximations, yet non-universal properties of the vortex liquid and of the first order transition can only be trusted qualitatively.
In the presence of non-zero stiffness $K$, the behavior of the $\theta$-field 
needs to be discussed on top of the given $\Psi$ solution. This leads to
three candidate phases. To understand this, first freeze $\Psi$ at $\Psi_2$ and study the field theory $S[\theta, \Psi_2]$. The effective fugacity $\mathcal Y_{\rm eff} = \mathcal Y e^{\Psi_2}$ is large, and the cosine term in the action cannot be treated perturbatively; thus in
the \emph{vortex liquid} $\theta$ is pinned, i.e., the system is a normal conductor. In contrast, when we freeze $\Psi$ at $\Psi_1 {\ll 1}$, the effective fugacity $\mathcal Y e^{\Psi_1}$ is small, the logarithm can be expanded and the effective sine-Gordon field theory $S[\theta, \Psi_1]$ is well-known to contain two phases:
If $K$ is small, the cosine term remains RG relevant even at small fugacity, i.e.
the \emph{vortex gas} discussed above survives to finite $K$. If $K$ is large enough, however, the cosine term becomes RG irrelevant and $\theta$ fluctuates freely, 
stabilizing the 
\emph{superfluid} (free vortices are absent, all vortices being bound in pairs).

This simple classification, based on locating the stationary points of the bare potential, is only a starting point, and it leaves two questions open. First: as $K$ is tuned while $\Psi$ stays pinned near $\Psi_1$, is the vortex-gas--superfluid transition governed by the same fixed-point structure as the ordinary BKT transition? Second: can the competition between the two minima $\Psi_1$ and $\Psi_2$ turn the liquid--gas/superfluid transition first order rather than continuous, and if so, in what range of parameters? Both questions are taken up below.

\section{The renormalization group analysis}
\label{sec: RG}
Having established the effective action $S[\theta,\Psi]$ governing the coupled
phase and chemical potential  sectors, we now analyze its stability
using a perturbative renormalization-group (RG) 
analysis near the fixed point of decoupled fields, i.e., expanding Eq.~\eqref{eq: sine-Gordon action} about $\mathcal Y = 0$. 

\subsection{General considerations}
We anticipate on the basis of $S_{\rm SR}[\Psi]$ that fluctuations of the chemical potential are gapped on length scales $L \gtrsim \xi_F$. Thus the RG contains multiple steps. First, we consider length scales 
 $\xi\lesssim L\lesssim\xi_F$ where both $\theta$ and $\Psi$ fields are approximately gapless (recall that the vortex core size $\xi$ is the UV cut-off of our theory). Unless $\mathcal Y$ has reached strong coupling within this first RG step, only the $\Psi$-field is pinned for $L \gtrsim \xi_F$. Fluctuations of $\theta$ are then incorporated in a second RG treatment, or (if the fluctuating chemical potential is pinned near $\Psi_2$) using the instanton analysis in the next section. Using this procedure we estimate the phase boundaries of Fig.~\ref{fig: phase diagram}.

We note that, for 
the RG treatment 
for intermediate $\xi \leq L \leq \xi_F$, the most general local action consistent with the symmetries of the
problem contains, in addition to the Gaussian terms, a tower of operators
of the form
$e^{\alpha\Psi}\cos(n\theta)$, $\alpha, n \in \mathbb N$, and their gradient-dressed counterparts. A subset of such operators --- which may be relevant --- is generated at higher orders of the RG. However,
rather than introducing all of these couplings explicitly here, we give
the complete symmetry-allowed action and its one-loop RG equations in
Appendix~\ref{ap: RG}. In particular, only the
sector with $\alpha+n$ even is generated (this parity is preserved under
RG, and hence sufficient to check on the bare couplings), while the additional gradient-dressed couplings
generated under RG remain parametrically subleading to the fugacity. The important point for the  
main text discussion (next subsections \ref{sec:SimpleAction}, \ref{sec:LeadingYRG}) is that the RG
flow generated from the microscopic fugacity $\mathcal Y$ is controlled and
reproduces the form of the action only to leading order in the fugacity.
Beyond this order the generated couplings scale differently from one another,
so that the full logarithmic form of the interaction in
Eq.~\eqref{eq: sine-Gordon action} is not preserved under RG. The
leading-order discussion of the main text is therefore limited to weak
$\mathcal Y$ (and should be supplemented by numerical integration of the
rather convoluted RG equations, see Appendix~\ref{ap: RG} Eq.~\eqref{eq: full RG system}). Within this
leading-order treatment, we obtain the renormalized BKT transition
temperature, an estimate for the first order transition line and identify below the parameter regime in which our analysis
applies. The renormalized boundaries of the phase diagram are discussed in
Sec.~\ref{sec: renormalized boundary} and the instanton analysis of
Sec.~\ref{sec: instantons} uses the bare potential, Eq.~\eqref{eq:VthetaPsi}.

\subsection{Approximate action to leading order in fugacity}
\label{sec:SimpleAction}
The effective action in the small $\mathcal Y$ limit and for $L<\xi_F$, which is subject to the RG treatment, follows from Eq.~\eqref{eq: sine-Gordon action}.
The short-range part takes the form
\begin{equation}
S_{\rm SR}[\Psi]=\int\frac{d^2p}{(2\pi)^2}\,\frac{\xi}{8\pi^2\kappa}\,
\big(\mathcal M^2\xi^{-2}+\mathbf p^2\big)^{3/2}\,|\Psi(\mathbf p)|^2 .
\label{eq: full theta Psi action}
\end{equation}
and the interaction term~\eqref{eq:IntTerm} we write as:
\begin{equation}
S_{\rm int}[\theta,\Psi]=-\frac{1}{\xi^2}\int d^2x\;\mathcal Y e^{\Psi(\mathbf x)}\cos\theta(\mathbf x).
\label{eq:IntTermApprox}
\end{equation}
The starting values of the dimensionless coupling constants are $\kappa(0) = \frac{E_0 \xi}{2\pi T \xi_F}$ and $\mathcal M(0) = \frac{\xi}{\xi_F}.$ This expression
follows from the Fourier transform of the short-ranged
Majorana-mediated interaction and is derived explicitly in
Appendix~\ref{ap: RG}; the
cubic dispersion for momenta $\vert p \vert \gtrsim 1/\xi_F$
originates
there from the Fourier transform of 
$V^{\rm ferm}$ in the regime $E_0 e^{- r/\xi_F} \gg T$ valid during the first stage of RG flow $\xi \lesssim r \lesssim \xi_F$. 

We briefly discuss the properties of the effective potential $ b
\Psi^2-\mathcal Y  e^\Psi {\cos(\theta)}$ (bare value $b = \mathcal M^3/[8 \pi^2\kappa]$) and discuss how the approximate Eq.~\eqref{eq:IntTermApprox} can be used to estimate the window, Eq.~\eqref{eq:WindowFor2Min}, of phase coexistence. First, we note that for homogeneous field configurations, the vortex concentration based on Eq.~\eqref{eq:IntTermApprox} is $n \sim \mathcal Y e^{\Psi}/\xi^2$ which must be less than $1/\xi^2$ to be internally consistent with the UV regularization imposing at most one vortex per area $\xi^2$. We next note that, contrary to the full logarithmic expression~\eqref{eq:IntTerm}, Eq.~\eqref{eq:IntTermApprox} is unbounded and always features a runaway minimum at $\theta = 0, \Psi \rightarrow \infty$. In addition, it contains a local minimum $\Psi_1$ and a maximum $\Psi_0$ if $b> b_< \simeq \mathcal Y e/2$. The window of coexisting physical solutions can be estimated using Eq.~\eqref{eq:IntTermApprox} by requiring that the density at the maximum is consistent with the UV regularization. 
This density is given by $\left .n \right \vert_{\rm max} \sim \mathcal Y e^{\Psi_0}/\xi^2$, where $\Psi_0 \simeq \ln\left (\frac{2 b}{\mathcal Y} \ln \left ( \frac{2b}{\mathcal Y}\right ) \right)$. It is physical only for $b < b_> \simeq 1/[2\ln(1/\mathcal Y)]$. Crucially, the window of two coexisting physical solutions obtained from Eq.~\eqref{eq:IntTermApprox} is identical to Eq.~\eqref{eq:WindowFor2Min}. The first order transition of two degenerate minima can be expected close to $b_>$, where the exponential term grows quickly.

\subsection{Leading order RG equations}
\label{sec:LeadingYRG}

The  
one-loop RG equations are
\begin{subequations}
\begin{align}
{\frac{dK}{d l}} & {= 0}, &
\dfrac{d {\kappa}}{dl}&={\kappa},\\
\frac{d {\mathcal M}}{dl}&={\mathcal M},
&
\frac{d\mathcal{Y}}{dl}
&=
\left(
2-\pi K+{\pi \kappa}
\right)\mathcal{Y}.
\end{align}

\label{eq: reduced RG system}
\end{subequations}
where $l=\ln(L/\xi)$ is the logarithmic RG scale.

The flow of ${\mathcal M}$ identifies $\xi_F$ as the scale at which the 
fluctuating chemical potential becomes massive:
\begin{equation}
{\mathcal M}(l)={\mathcal M}(0)e^l,
\quad {\mathcal M}(l_{\mathcal M})\sim 1
{\Leftrightarrow}
L_{\mathcal M}=\xi e^{l_{\mathcal M}}\sim\xi_F.
\label{eq: Lm definition}
\end{equation}
Thus, $\xi_F$ is not an additional arbitrary scale introduced into the RG
analysis; it is precisely the scale at which the mass term becomes of
order unity and the long-distance 
fluctuations of the chemical potential cease to be
important.

The leading fugacity (i.e., the coefficient of the leading term in the expansion of the logarithm), on the other hand, reaches strong coupling when
$\mathcal{Y}
(l_{\mathcal Y})\sim1$, defining a second characteristic scale
$L_{\mathcal Y}=\xi e^{l_{\mathcal Y}}$. Since
\begin{equation}
{\kappa(l) = \kappa(0) e^l},
\end{equation}
the fugacity equation can be integrated explicitly,
\begin{equation}
\mathcal{Y}(l)
=
\mathcal{Y}(0)
\exp\left[
\left(2-\pi K\right)l
+ {\pi \kappa(0)
}
(e^l-1) 
\right]. \label{eq:YRenorm}
\end{equation}
Consequently, $L_{\mathcal Y}$ is determined by
\begin{equation}
1=
\mathcal{Y}(0)
\exp\left[
\left(2-\pi K\right)
\ln\frac{L_{\mathcal Y}}{\xi}
+\pi \kappa(0) (L_{\mathcal Y}/\xi-1)
\right].
\label{eq: LY equation}
\end{equation}
\subsection{Crossover to semiclassical field configurations}
{The couplings $K$ and $\mathcal{Y}(0)$ entering the conditions below depend on temperature. In terms of the temperature-independent microscopic energy scales $J$ and $E_{\rm core}$ they read
\begin{equation}
K = \frac{J}{T}, \qquad
\mathcal{Y}(0) = 2 e^{\mu/T} = 2\sqrt{2}\, e^{-E_{\rm core}/T}.
\label{eq:BareCouplingsT}
\end{equation}
Here $J$ is the physical, dimensionful stiffness set by the superfluid density. In the following, we state each boundary of the phase diagram first in terms of $K$ and $\mathcal{Y}(0)$ and then, using Eq.~\eqref{eq:BareCouplingsT}, in terms of $J$, $E_0$, $E_{\rm core}$ and $T$.}

The relative ordering of $L_{\mathcal Y}$ and $L_{\mathcal M}\sim\xi_F$ determines which
degree of freedom ($\theta$ or $\Psi$) becomes strongly coupled first. The boundary between the two regimes can be obtained by evaluating the
fugacity flow at the scale $L_{\mathcal M}\simeq\xi_F$. Using Eq.~\eqref{eq: LY equation},
the condition $L_{\mathcal Y}=L_{\mathcal M}$ is
\begin{align}
&\mathcal{Y}(0)
\exp\left[
\left(2-\pi K\right)
\ln\frac{\xi_F}{\xi}
+ {\frac{E_0 }{2T}}
\left (1-\frac{\xi}{\xi_F} \right )
\right]
=1. 
\label{eq: regime boundary}
\end{align}
Thus, {to leading order in $\xi/\xi_F \ll 1$,} the regime $L_{\mathcal M}<L_{\mathcal Y}$, in which the chemical
potential sector becomes massive before the vortex fugacity reaches
strong coupling, occurs for
{
\begin{subequations}
\label{eq:LMLYCond}
\begin{align}
\frac{E_0}{T} &< - 2\ln \left [\mathcal Y(0)\left (\frac{\xi_F}{\xi} \right)^{2-\pi K} \right ], \label{eq:LMLYCond_a}\\
\frac{E_0}{T} &< \frac{2E_{\rm core} + 2\pi J \ln(\xi_F/\xi)}{T} - 2\ln\!\left[2\sqrt{2}\left(\frac{\xi_F}{\xi}\right)^{2}\right]. \label{eq:LMLYCond_b}
\end{align}
\end{subequations}}
The parameter regime considered below is chosen such that this condition is satisfied over the relevant
range of $E_0/T$.

We now discuss the two regimes separately. If
\begin{equation}
L_{\mathcal Y}<L_{\mathcal M},
\label{eq: regime2}
\end{equation}
the vortex fugacity reaches strong coupling while the auxiliary field is
still effectively light. In this case the disorder field $\theta$ is already
locked on scales shorter than $\xi_F$, and the system is always a normal conductor.
The remaining chemical potential sector can then be
viewed as an effective theory in the background of the strongly coupled sector of the disorder field with pinned $\theta$.
Depending on the parameters, this regime contains the
distinct normal-state phases associated with the vortex liquid--gas
transition. A detailed analysis of this transition is beyond the scope
of the present work, since our focus is on the regime in which the
Majorana-mediated interaction modifies the vortex unbinding physics
rather than on the structure of the normal state itself.

In the opposite regime
\begin{equation}
L_{\mathcal M}<L_{\mathcal Y},
\label{eq: regime1}
\end{equation}
the chemical potential sector becomes massive while the vortex fugacity, Eq.~\eqref{eq:YRenorm}, is still small.
At this point the $\Psi$ field can be integrated out, leaving an
effective theory of the phase and vortices with $\xi_F$ as its ultraviolet
cutoff. 

\subsection{Renormalized BKT transition temperature}
\label{sec:Effective SG}

We here use the renormalization group calculation to estimate corrections to the BKT transition temperature. To this end, we assume that $\Psi$ is pinned at the minimum $\Psi_1$, which is close to zero (note that in this regime $L_{\mathcal M} < L_{\mathcal Y}$). The coupling constants at the end of the first RG step, in particular the fugacity $\mathcal Y\equiv \mathcal Y(l_{\mathcal M})$
then serve as input for the subsequent sine--Gordon renormalization group study, and the transition occurs when
\begin{equation}
    \mathcal Y(l_{\mathcal M}) = K- 2/\pi. 
\end{equation}
Solving this equation in the regime of interest defined by Eq.~\eqref{eq:LMLYCond} leads to
{\begin{subequations}
\label{eq:BKTRENORM}
\begin{align}
K &\simeq \frac{2}{\pi} + \mathcal{Y}(0)\, e^{\frac{E_0}{2T}}, \label{eq:BKTRENORM_a}
\\
\frac{J}{T_{\rm BKT}} &\simeq \frac{2}{\pi} + 2\sqrt{2}\, \exp\!\left[-\frac{E_{\rm core} - E_0/2}{T_{\rm BKT}}\right].\label{eq:BKTRENORM_b}
\end{align}
\end{subequations}}
This concludes the derivation of the BKT transition line depicted in Fig.~\ref{fig: phase diagram inset}.

\subsection{Renormalized boundary of first-order transition}
\label{sec: renormalized boundary}

We may finally estimate the regime of coexisting solutions using the same arguments based on Eq.~\eqref{eq:IntTermApprox}, but using renormalized coupling constants in the regime $L_{\mathcal M} \sim \xi_F < L_{\mathcal Y}$. The renormalized quadratic coefficient is $b(l_{\mathcal M}) = b(0) \xi_F^2/\xi^2$, the renormalized maximum occurs at $\Psi_0 \simeq \ln\!\Big(\frac{2b(l_{\mathcal M})}{\mathcal Y(l_{\mathcal M})}\ln\frac{2b(l_{\mathcal M})}{\mathcal Y(l_{\mathcal M})}\Big)$ and the vortex density including leading fluctuation effects is thus at the maximum
\begin{equation}
    \left. n \right \vert_{\rm max} = \mathcal Y (l_{\mathcal M}) e^{\Psi_0}/\xi_F^2. 
\end{equation}
For consistency it must be less than $\xi_F^{-2}$, i.e.\ $\mathcal Y(l_{\mathcal M})e^{\Psi_0}<1$: the RG equations used to obtain $\mathcal Y(l_{\mathcal M})$ are of leading order in the fugacity and remain controlled only as long as the number of vortices per area of the running cutoff, here $\xi_F^2$, is small. 
Using this we find the renormalized coexistence region $b(0) \in [b_<, b_>]$, where
\begin{subequations}
\label{eq:RenWindow}
\begin{align}
    b_< &\simeq \frac{e \mathcal Y(0)}{2} e^{- \pi K \ln(\xi_F/\xi) + \pi \kappa(0) (\xi_F/\xi - 1)},\\
    b_> &\simeq \frac{(\xi/\xi_F)^2}{- 2 \ln\!\big(\mathcal Y(0)\,(\xi_F/\xi)^2\, e^{- \pi K \ln(\xi_F/\xi) + \pi \kappa(0) (\xi_F/\xi - 1)}\big)}.
\end{align}
\end{subequations}
Again, we expect the position of the first order transition close to $b_>$. In the absence of renormalizations, e.g. setting $\xi_F = \xi$, this expression reproduces Eq.~\eqref{eq:WindowFor2Min}.

Inserting the bare values of $b$ and $\kappa$ into Eqs.~\eqref{eq:RenWindow} and solving for $K$, we find two coexisting solutions for $K_<<K<K_>$, where
\begin{subequations}
\label{eq:Kwindow}
\begin{align}
K_<&=\frac{2}{\pi}+\frac{\ln\!\big(2\pi e\,\mathcal Y(0)\frac{E_0}{T}\big)+\frac{E_0}{2T}\big(1-\frac{\xi}{\xi_F}\big)}{\pi\ln(\xi_F/\xi)},\\
K_>&=\frac{2}{\pi}+\frac{2\pi\frac{E_0}{T}+\frac{E_0}{2T}\big(1-\frac{\xi}{\xi_F}\big)-\ln\frac{1}{\mathcal Y(0)}}{\pi\ln(\xi_F/\xi)}.
\end{align}
\end{subequations}
These boundaries are shown in Fig.~\ref{fig: phase diagram inset} and the first-order transition is expected close to $K_>$.
To locate it more precisely, we note that the renormalized window~\eqref{eq:RenWindow} follows from Eq.~\eqref{eq:WindowFor2Min} by the replacement $b\to b(l_{\mathcal M})$, $\mathcal Y\to\mathcal Y(l_{\mathcal M})$.
Applying the same replacement to the degeneracy condition~\eqref{eq:FirstOrder}, we estimate the first-order line as $b(l_{\mathcal M})=\frac{1/4}{\ln(1/\mathcal Y(l_{\mathcal M}))}$.
This gives
{\begin{subequations}
\label{eq:K1st}
\begin{align}
K_{\rm 1st} &= \frac{2}{\pi} + \frac{\pi\frac{E_0}{T} + \frac{E_0}{2T}\big(1 - \frac{\xi}{\xi_F}\big) - \ln\frac{1}{\mathcal{Y}(0)}}{\pi \ln(\xi_F/\xi)},\label{eq:K1st_a} \\
\frac{J}{T_{\rm 1st}} &= \frac{2}{\pi} + \frac{\pi\frac{E_0}{T_{\rm 1st}} + \frac{E_0}{2T_{\rm 1st}}\big(1 - \frac{\xi}{\xi_F}\big) - \frac{E_{\rm core}}{T_{\rm 1st}} + \ln\big(2\sqrt{2}\big)}{\pi \ln(\xi_F/\xi)},\label{eq:K1st_b}
\end{align}
\end{subequations}}
which indeed lies inside the window, between $K_<$ and $K_>$.

\subsection{Discussion of the phase diagram}
\label{sec:PhaseDiagDiscussion}

The results of this and the previous section can be used to provide details for Fig.~\ref{fig: phase diagram inset}.
As mentioned in the caption, the shading indicates the global minimum of the free energy: superfluid (purple) and vortex gas (gray) correspond to $\Psi_1$ and are separated by the BKT line, Eq.~\eqref{eq:BKTRENORM_a} (red), the vortex liquid (blue) corresponds to $\Psi_2$.
{The black line is the mean-field estimate of the first-order transition, Eq.~\eqref{eq:K1st_a}.}
The hatched orange region is the window of two coexisting minima, $K_<<K<K_>$, Eq.~\eqref{eq:Kwindow}.
Beyond its intersection with the first-order line, the BKT line (dashed) refers to the metastable $\Psi_1$ branch.
The dash-dotted line marks $L_{\mathcal M}=L_{\mathcal Y}$, Eq.~\eqref{eq:LMLYCond_a}; our analysis applies to the left of it, while to the right ($L_{\mathcal Y}<L_{\mathcal M}$) the vortex fugacity reaches strong coupling before the chemical-potential sector becomes massive.

\begin{figure*}
\includegraphics[width=\textwidth]{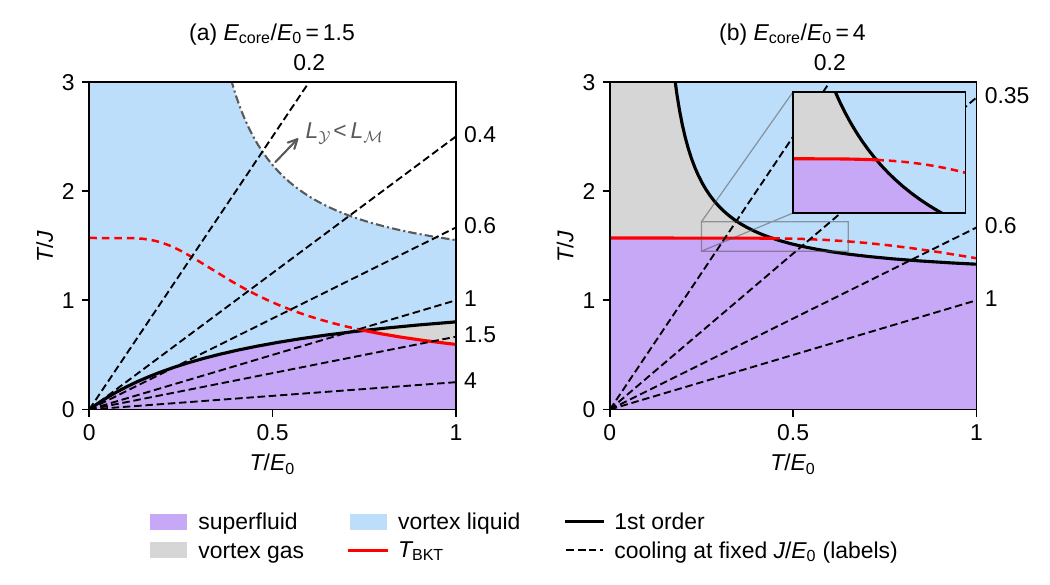}
\caption{\justifying {Phase diagram at temperature-independent $J$, $E_0$ and $E_{\rm core}$, with $\mathcal{Y}(0)=2\sqrt{2}\,e^{-E_{\rm core}/T}$ and $\xi_F/\xi=5$, for (a) $E_{\rm core}/E_0=1.5$ and (b) $E_{\rm core}/E_0=4$, i.e.~below and above $E_{\rm core}/E_0=\pi+\tfrac{1}{2}\big(1-\tfrac{\xi}{\xi_F}\big)\approx 3.54$. Red line: BKT transition, Eq.~\eqref{eq:BKTRENORM_b}. Black line: first-order transition, Eq.~\eqref{eq:K1st_b}. The condition $L_{\mathcal{M}}<L_{\mathcal{Y}}$, Eq.~\eqref{eq:LMLYCond_b}, is satisfied throughout the plotted range of both panels. Dashed lines are cooling trajectories at fixed $J/E_0$ (values indicated at the frame), directed towards the origin. Inset in (b): termination of the BKT line on the first-order line. Note the different scales of the vertical axes.}}
\label{fig:cooling}
\end{figure*}

In an experiment in which the temperature is varied at fixed microscopic energy scales {$J$, $E_0$ and $E_{\rm core}$, the fugacity changes together with $E_0/T$ and $K=J/T$, see Eq.~\eqref{eq:BareCouplingsT}.
To see which sequences of phases are then possible, we show in Fig.~\ref{fig:cooling} the phase diagram obtained from Eqs.~\eqref{eq:LMLYCond_b}, \eqref{eq:BKTRENORM_b} and \eqref{eq:K1st_b} in the plane $(T/E_0,T/J)$ for two values of $E_{\rm core}/E_0$, one below and one above the value $E_{\rm core}/E_0=\pi+\tfrac{1}{2}\big(1-\tfrac{\xi}{\xi_F}\big)$ at which the layout of the phase diagram changes, see below. In this representation cooling at fixed $J/E_0$ corresponds to a straight line towards the origin.}

Two features are generic. First, the vortex liquid always lies on the high-temperature side of the first-order line, so that a cooling trajectory crosses this line at most once. {A solution $T_{\rm 1st}>0$ exists only if
\begin{equation}
E_{\rm core} + \pi J \ln\frac{\xi_F}{\xi} > E_0\Big[\pi + \tfrac{1}{2}\big(1-\tfrac{\xi}{\xi_F}\big)\Big].
\label{eq:EcoreMin}
\end{equation}
Otherwise the Majorana hybridization gain, given by the right-hand side, exceeds the core and logarithmic energy of a vortex on the scale $\xi_F$, and the system remains a vortex liquid down to the lowest temperatures. 
Second, upon heating the superfluid is destroyed at the lower of the two temperatures $T_{\rm 1st}$ and $T_{\rm BKT}$. If $T_{\rm 1st}<T_{\rm BKT}$, the transition is of first order, directly into the vortex liquid Fig.~\ref{fig:cooling}(a), (b). If $T_{\rm BKT}<T_{\rm 1st}$, the ordinary BKT transition into the vortex gas takes place first, so this requires a core energy large compared with the hybridization gain Fig.~\ref{fig:cooling}(b). The two temperatures coincide where the BKT line terminates on the first-order line. For small core energies the BKT line reappears at large $J/E_0$ Fig.~\ref{fig:cooling}(a), but there the renormalized fugacity $\mathcal{Y}(l_{\mathcal{M}})$ is of order unity and our leading-order treatment is only qualitative.}

\section{Instanton  calculation of the decay rate}
\label{sec: instantons}

We now turn to instanton (bounce) calculus to compute the decay rate of
the metastable phase, in the regime $L_{\mathcal M}<L_{\mathcal Y}$, where the chemical potential sector
becomes massive, while the fugacity remains small. This decay rate is
directly connected to the first-order character of the transition: the
potential $V(\theta,\Psi)$ possesses two nearly degenerate minima
separated by a barrier, and the false-vacuum minimum decays into the true
vacuum via nucleation of bubbles --- instantons/bounces --- whose action
controls the {nucleation rate. The decay is thermally activated and the bounce is the critical nucleus of Langer's theory of nucleotion~\cite{langer1969statistical,langer2000theory} and the calculation is formally equivalent to the false-vacuum-decay framework of Coleman~\cite{coleman1977fate} and
Callan--Coleman~\cite{callan1977fate}.}
Before presenting the calculation,
it is worth summarizing the logic in one paragraph: one identifies the
(nearly degenerate) minima and the saddle point of the potential,
constructs  
the isotropic bounce solution interpolating between
them, evaluates the Euclidean action of this solution (the bounce action
$B$), and combines it with the Gaussian fluctuation determinant about the
bounce (relative to the same determinant evaluated in the false vacuum)
to obtain the decay rate per unit area, $\Gamma\sim\mathcal{X}e^{-B}$, where the prefactor $\mathcal{X}$, which collects the zero-mode measure and the ratio of determinants, has the dimension of $T/\xi^2$. As discussed below, and contrary to usual instanton calculus, the inside of the bounce (i.e. the nucleation bubble) is characterized by a quasi-gapless field whose contribution parametrically 
enhances the prefactor $\mathcal{X}$ of the decay rate. 

The second minimum $\Psi_2$ is a property of the full logarithmic interaction, which, as discussed above, is not preserved under RG beyond leading order in the fugacity. We therefore construct the bounce from the bare action, Eq.~\eqref{eq: sine-Gordon action}, with bare couplings (this works well when the fluctuation corrections discussed in the previous section are small). We expand the kernel of Eq.~\eqref{eq: full theta Psi action} to order $\mathbf p^2$. The zeroth order is the mass term $\frac{\mathcal M^3}{8\pi^2\kappa}\Psi^2$, i.e.\ $b\Psi^2$ of the bare potential, and an ordinary gradient term for $\Psi$:
\begin{equation}
S[\theta,\Psi]=\int d^2x\left[\frac{1}{8\pi^2K}(\nabla\theta)^2+\frac{3\mathcal M}{16\pi^2\kappa}(\nabla\Psi)^2+\frac{V(\theta,\Psi)}{\xi^2}\right],
\label{eq: instanton effective action}
\end{equation}
with the bare potential $V(\theta,\Psi)$, Eq.~\eqref{eq:VthetaPsi}.
The kernel of Eq.~\eqref{eq: full theta Psi action} assumes $E_0e^{-r/\xi_F}\gg T$.
Hence the conditions Eqs.~\eqref{eq:WindowFor2Min} and \eqref{eq:FirstOrder} apply directly to the potential of Eq.~\eqref{eq: instanton effective action}.
The gradient expansion is parametrically justified only if $\Psi$ varies across the bubble wall on a length large compared with $\xi_F$, and the thin-wall approximation requires this length to be small compared with the bubble radius $R_B$.

We use the stationary points $\Psi_1<\Psi_0<\Psi_2$ of $V(0,\Psi)$, Fig.~\ref{fig: potential landscape}, and work on the side of the first-order line where $\Psi_1$ is the true vacuum and $\Psi_2$ the metastable one.

The thin-wall bounce construction requires near-degenerate minima, i.e.\ a splitting $\epsilon=[V(0,\Psi_2)-V(0,\Psi_1)]/\xi^2$ small compared with the barrier height $\Delta V=[V(0,\Psi_0)-V(0,\Psi_2)]/\xi^2$. Close to the first-order line, Eq.~\eqref{eq:FirstOrder}, which is exactly $4b\ln(1/\mathcal Y)=1$, one finds (App.~\ref{ap: wall tension})
\begin{equation}
\epsilon\simeq\frac{1}{\xi^2}\Big[\ln\frac1{\mathcal Y}-\frac1{4b}\Big],\qquad
\Delta V\simeq\frac{1}{4\xi^2}\ln\frac1{\mathcal Y}.
\label{eq: eps DeltaV}
\end{equation}
The splitting $\epsilon$ thus measures the distance from the first-order line. Since $\ln(1/\mathcal Y)\gg1$, the thin-wall condition $\epsilon\ll\Delta V$  holds for all $\epsilon\xi^2\ll\frac14\ln(1/\mathcal Y)$, including $\epsilon\xi^2=\mathcal O(1)$.

The bounce is a radially symmetric configuration of $\Psi$ at $\theta=0$: a bubble of the true vacuum $\Psi_1$ of radius $R_B$ embedded in the false vacuum $\Psi_2$. In the thin-wall limit $\epsilon\ll\Delta V$ its action is $B(R)=2\pi R\sigma_\Psi-\pi R^2\epsilon$, which is extremal at $R_B=\sigma_\Psi/\epsilon$, giving $B=\pi\sigma_\Psi^2/\epsilon$. The wall tension follows from the first integral for a flat domain wall,
\begin{equation}
\sigma_\Psi=\frac{1}{\xi}\sqrt{\frac{3\mathcal M}{4\pi^2\kappa}}\int_{\Psi_\star}^{\Psi_2}d\Psi\,\sqrt{V(0,\Psi)-V(0,\Psi_2)},
\label{eq: wall tension integral}
\end{equation}
where $\Psi_\star$ is the turning point, $V(0,\Psi_\star)=V(0,\Psi_2)$.
To zeroth order in $\epsilon$ the integral can be evaluated on the first-order line, where $V(0,\Psi)$ consists of two parabolic wells up to relative corrections $\mathcal O(1/\ln^2 (1/\mathcal Y))$ (App.~\ref{ap: wall tension}), so that
\begin{align}
\sigma_\Psi&\simeq\frac{\sqrt6\,\mathcal M^2}{8\pi^2\kappa\,\xi}\,\ln^2\frac1{\mathcal Y},\label{eq: wall tension}\\
B&\simeq\frac{3\mathcal M^4}{32\pi^3\kappa^2}\,\frac{\ln^4(1/\mathcal Y)}{\epsilon\xi^2},\label{eq: bounce action}\\
R_B&=\frac{\sigma_\Psi}{\epsilon}\simeq\frac{\sqrt6\,\mathcal M^2}{8\pi^2\kappa}\,\frac{\ln^2(1/\mathcal Y)}{\epsilon\,\xi}.\label{eq: bounce radius}
\end{align}
The wall approaches both vacua exponentially on the scale 
$\xi_F$, so its thickness, $\ell_{\rm w}(\delta)\simeq\sqrt6\,\xi_F\ln[\ln(1/\mathcal Y)/\delta]$, depends logarithmically on the accuracy $\delta$ to which the vacua are reached (App.~\ref{ap: scales}). The wall tension, however, is accumulated within a distance of order $\xi_F$ of the wall center, 
hence the gradient expansion is controlled  
up to factors of order unity, and $R_B/(\sqrt6\,\xi_F)=\Delta V/\epsilon$, so $R_B\gg \xi_F$ is precisely the thin-wall condition $\epsilon\ll\Delta V$. {Accordingly, the numerical coefficient of $\sigma_\Psi$, and hence of $B$, is determined only up to a factor of order unity, while the parametric dependence of $B$ and the conclusion $B\gg1$ are unaffected.}

\begin{figure}
    \centering
    \includegraphics[width=0.4\textwidth]{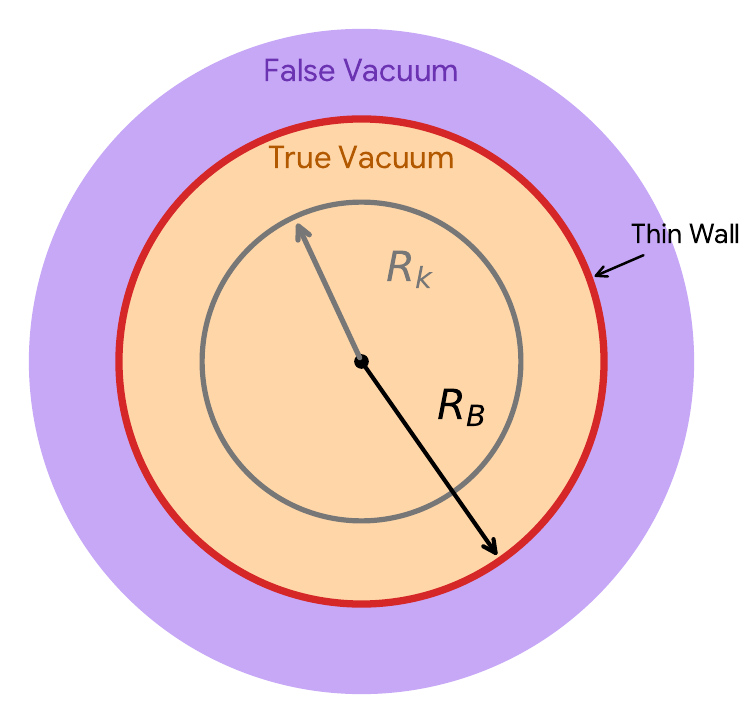}
    \caption{\justifying Bounce configuration. A bubble of radius $R_B$ of the true vacuum $\Psi_1$, in which $\theta$ approximately fluctuates freely, is embedded in the false vacuum $\Psi_2$, in which $\theta$ is pinned. The thin wall imposes the Dirichlet condition $\theta=0$ at $r=R_B$. $R_k$ denotes the width of a kink of $\theta$ inside the bubble.}
    \label{fig: coleman-bounce}
\end{figure}

The bounce itself does not involve $\theta$, the phase enters only through the fluctuations about the bounce, i.e.\ through the prefactor of the decay rate. Outside the bubble, in the false vacuum $\Psi_2$, the effective fugacity $\mathcal Ye^{\Psi_2}$ is large and $\theta$ is pinned at $\theta=0$. Inside the bubble, in the true vacuum $\Psi_1$, the effective fugacity $\mathcal Ye^{\Psi_1}\simeq\mathcal Y$ is small. Since $\theta$ is not compact, a bounce retaining the fugacity term inside must be accompanied by kink--antikink pairs in $\theta$. For $K>2/\pi$ the cosine is irrelevant and these pairs proliferate, so that $\theta$ is approximately a free field inside the bubble, and the pinning outside imposes a Dirichlet condition on the bubble wall, see Fig.~\ref{fig: coleman-bounce}:
\begin{equation}
S^{(2)}[\theta]\simeq\frac1{8\pi^2K}\int_{|\mathbf x|<R_B}d^2x\,(\nabla\theta)^2,
\qquad
\theta|_{|\mathbf x|=R_B}=0.
\label{eq: theta free action}
\end{equation}
The Gaussian integral over $\theta$ with the action~\eqref{eq: theta free action} gives (App.~\ref{ap: instanton})
\begin{equation}
Z_\theta^{\rm free}(R_B)=e^{Z_D'(0)/2}(4\pi^2K)^{1/12}\Big(\frac{R_B}{\xi}\Big)^{1/6},
\label{eq: Z theta free}
\end{equation}
where $Z_D(s)=\sum_{m,n}j_{m,n}^{-2s}$ is the spectral $\zeta$-function of the Dirichlet disk and $j_{m,n}$ are the zeros of the Bessel function $J_m$.

Combining these ingredients yields the decay rate per unit area, $\Gamma$, in the regime $L_{\mathcal M}<L_{\mathcal Y}$ and under the near-degeneracy condition; the full expression is given in Appendix~\ref{ap: instanton}. The static calculation fixes the exponent and the prefactor up to the time scale of the decay. The latter is set by the growth rate of the unstable radial mode of the bubble, which depends on the dynamics and is not fixed by the action; we estimate it as $T$. To leading exponential accuracy for $E_0/T\gg1$ and $\xi_F\gg\xi$, and with the numerical coefficient in the exponent subject to the uncertainty of the gradient expansion discussed above,
\begin{equation}
\Gamma\sim\frac{T}{\xi^2}\exp\left[-\frac{3\pi^3}{8}\left(\frac{E_0}{T}\right)^2\left(\frac{\xi_F}{\xi}\right)^6\frac{1}{\epsilon\xi^2}\right].
\label{eq: decay rate result}
\end{equation}
The rate depends on the distance from the first-order line through $\epsilon$. On the line, $\epsilon\to0$, the bounce action diverges. Throughout the thin-wall regime $\epsilon\ll\Delta V$ the bounce action remains very large, $B\gg1$, so that the metastable phase is stable on any experimental time scale. When the first-order line is crossed, the system therefore remains in the metastable phase until the boundary of the coexistence window, Eq.~\eqref{eq:Kwindow}, is approached, which leads to hysteresis.

\section{Discussion and Outlook}
In this work, we have asked whether a two-dimensional topological superconductor can host isolated, well-separated Majorana zero modes in its vortex cores while simultaneously supporting genuine BKT physics --- namely, the hierarchy $E_0>T_{BKT}$ together with $\xi_F\gg\xi$ set out in Sec.~\ref{sec: two vortex problem}. Starting from the two-vortex problem, we showed that Majorana hybridization between nearby vortex cores modifies the standard logarithmic vortex-antivortex interaction at short distances, generating an emergent local minimum in the pair free energy.
We identified a microscopic platform realizing this hierarchy, an array of Josephson-coupled $p+ip$ superconducting islands, and found that it can satisfy the required ordering of scales in an accessible parameter window.

Building on this result, we constructed an effective field theory for the resulting dilute vortex gas with Majorana-mediated interactions: a sine-Gordon-type action coupling the disorder field $\theta$, dual to the superconducting phase, to a Hubbard--Stratonovich field $\Psi$ that decouples the short-range interaction. The RG analysis identifies the regime relevant to our hierarchy, in which $\Psi$ becomes massive, while the vortex fugacity remains small. In this regime, the competition between the short-distance Majorana-mediated interaction and the long-distance logarithmic vortex interaction produces a first-order transition. Instanton methods show that the decay of the metastable state proceeds with an exponentially suppressed rate $\Gamma$. The phase diagram of the system follows from the potential $V(\theta,\Psi)$ evaluated with the couplings renormalized down to scale $\xi_F$: the conditions for two minima yield the coexistence window, their degeneracy yields the first-order line, and the renormalized BKT criterion separates the superfluid from the vortex gas. {A controlled description of the dense vortex phase, beyond the pairwise treatment of the Majorana hybridization, is left for future work.}

We conclude with a discussion of experimental consequences. First, the superconducting transition itself can be of first order: upon heating at fixed $J$, $E_0$ and $E_{\rm core}$, the superfluid gives way directly to the vortex liquid, Fig.~\ref{fig:cooling}. We emphasize that the fixed-fugacity diagram, Fig.~\ref{fig: phase diagram inset}, might suggest a reentrant normal state upon cooling, reminiscent of the Pomeranchuk effect~\cite{Pomeranchuk1950,Richardson1997Pomeranchuk}. This is not the case: once the temperature dependence of the fugacity is taken into account, the vortex liquid is always the high-temperature phase. 
 
Second, the first-order transition implies hysteresis. Since $\Gamma$ is exponentially small throughout the thin-wall regime, the width of the hysteresis loop is set by the coexistence window and is essentially independent of the sweep rate. A sweep-rate dependence, as in switching-current measurements on Josephson junctions~\cite{devoret1985measurements, martinis1987experimental, kurkijarvi1972intrinsic}, can arise only close to the limits of metastability, where the barrier becomes small and a thick-wall treatment beyond the present work is required.

A possible platform in this context is provided by recent reports of chiral 2D superconductivity~\cite{han2025signatures}, in which the coherence length and interparticle spacing are comparable --- the regime in which the Majorana-mediated vortex interaction considered here becomes relevant. 
The observation of such
first-order phenomenology
at the superconducting transition would provide evidence for subgap (Majorana) vortex-core states, distinguishing this scenario from a conventional, non-topological vortex gas.

\section*{AI Disclosure}
In the later stages of the project, the authors used Claude (Anthropic) Opus 5  for assistance with code for numerical calculations, for checking calculations, and for editing the manuscript, and take full responsibility for the content.

\acknowledgements

It is a pleasure to thank  M. Feigel'man and J. Schmalian for discussions and K. Chang, I. Esterlis, H. Guo, D. Jin, for a collaboration on a related topic. E.J.K. further acknowledges exploratory discussions over the past decade with several colleagues, notably with R.~Thomale and P.~Hosur, on the general question of Majorana modes and BKT physics, although the possibility of a first-order transition was not anticipated in those discussions.
Support for this research was provided by the Office of the Vice Chancellor for Research
and Graduate Education at the University of Wisconsin–Madison with funding from the Wisconsin Alumni
Research Foundation (E.J.K.). This work was performed in part at the Aspen Center for Physics, which is
supported by National Science Foundation grant PHY-
2210452 (E.J.K.). V.M. acknowledges funding from the Public Scholarship, Development, Disability and Maintenance Fund of the Republic of Slovenia through the 368th Public Call for scholarships for doctoral studies of foreign citizens in the field of Physics – Quantum Technologies (contract No. 11011-2/2025).

\appendix
\section{Josephson junction array}
\label{ap: JJ}
Here we give the details of the derivation of the effective tight-binding Hamiltonian for the boundary Majorana modes, and of the resulting bound $-2<M<0$ on the hybridization $M$ used above.

We start from the near-zero-energy boundary eigenmodes of a single island of radius $R$ (disk radius), coherence length $\xi$, and Fermi momentum $k_F$, labeled by an angular-momentum index $m\in\mathbb Z$ (shifted by one relative to the main text, so that the near-zero pair $m=-1,0$ of the main text corresponds to $m=0,1$ here),
\begin{align}
u_m(r) &= \frac{\mathcal N}{\sqrt r}\, e^{-(R-r)/\xi}\, e^{ik_F r}\, e^{im\theta}, \label{eq: u mode} \\
v_m(r) &= -\frac{i\mathcal N}{\sqrt r}\, e^{-(R-r)/\xi}\, e^{ik_F r}\, e^{i(m-1)\theta}, \label{eq: v mode}
\end{align}
where $(r,\theta)$ are polar coordinates measured from the center of the island and $\mathcal N$ is a normalization constant. We label the islands by their discrete positions $\v r$ and denote the continuum coordinate by $\v x$. The electron field operator on the island at $\v r$ is expanded as $\psi(\v x) = \sum_m u_m(\v x-\v r)\, c_m(\v r) + v_m^*(\v x-\v r)\, c_m^\dagger(\v r)$. For a point contact connecting the islands at $\v r,\v r'$, located at angles $\theta_s,\theta_s'$ on their respective boundaries, with tunneling amplitude $t$,
\begin{align}
H_{1\,\rm link} = -t\, \psi^\dagger(\v r' + R\hat e_{s}')\, \psi(\v r + R\hat e_{s}) + {\rm h.c.}
\end{align}
Inserting the mode expansion and relabeling $m\to 1-m,\ n\to 1-n$ so that every term shares the same angular factor $e^{i(m\theta_s-n\theta_s')}$, then passing to the Majorana basis $c_n = \tfrac12(\gamma_n+i\gamma_n')$ with the redundancy $\gamma_n=\gamma_{1-n}$, $\gamma_n'=-\gamma_{1-n}'$, gives
\begin{widetext}
\begin{align}
H_{1\,\rm link} = -\frac{t\mathcal N^2}{2R}\big(1+\sin 2k_F R\big) \sum_{m,n} e^{i(m\theta_s - n\theta_s')} \Big[
\gamma_n(\v r')\gamma_m(\v r) + i\gamma_n(\v r')\gamma_m'(\v r)
- i\gamma_n'(\v r')\gamma_m(\v r) + \gamma_n'(\v r')\gamma_m'(\v r)\Big] + {\rm h.c.}
\label{eq: H1link}
\end{align}
\end{widetext}
With contacts along the four orthogonal directions $\theta_1=0,\theta_2=\tfrac\pi2,\theta_3=\pi,\theta_4=\tfrac{3\pi}2$, Eq.~\eqref{eq: H1link} splits into a bond along $\hat x$ and a bond along $\hat y$,
\begin{align}
H_x &= -\frac{i\tilde t}{2} \sum_{m,n} (-1)^m \Big[ \gamma_n(\v r')\gamma_m'(\v r) - \gamma_n'(\v r')\gamma_m(\v r)\Big],
\end{align}
\begin{align}
H_y &= -\frac{i\tilde t}{2} \sum_{m,n} \Big\{ \sin\!\big[\tfrac{\pi}{2}(3m-n)\big] \notag \\
&\qquad \times \big[\gamma_n(\v r')\gamma_m(\v r) + \gamma_n'(\v r')\gamma_m'(\v r)\big] \notag \\
&\quad + \cos\!\big[\tfrac{\pi}{2}(3m-n)\big] \notag \\
&\qquad \times \big[\gamma_n(\v r')\gamma_m'(\v r) - \gamma_n'(\v r')\gamma_m(\v r)\big] \Big\},
\end{align}
with $\tilde t \equiv \tfrac{2t\mathcal N^2}{R}(1+\sin 2k_F R)$. We now restrict the sums to the near-zero-energy modes, $m,n\in\{0,1\}$, and use the redundancy $\gamma_1=\gamma_0$, $\gamma_1'=-\gamma_0'$ to express everything through the single Majorana pair $\gamma_0,\gamma_0'$. Writing the island positions as $\v r = la\,\hat x + ja\,\hat y$ (with lattice spacing $a$, set to $a=1$), the sum over bonds gives the real-space Hamiltonian for the boundary Majorana pair $\gamma_0,\gamma_0'$ on the resulting square lattice,
\begin{widetext}
\begin{align}
H = -i\tilde t \sum_{l,j} \Big[
2\,\gamma_0({l,j})\gamma_0'({l+1,j})
+ \gamma_0({l,j})\gamma_0'({l,j+1})
- \gamma_0'({l,j})\gamma_0({l,j+1})
- \gamma_0({l,j})\gamma_0({l,j+1})
+ \gamma_0'({l,j})\gamma_0'({l,j+1})
\Big].
\label{eq: Hinreal ap}
\end{align}
\end{widetext}
Fourier transforming as $\gamma_0(l,j) = \tfrac{1}{\sqrt N}\sum_{\mathbf k} e^{i(k_xla+k_yja)}\gamma_0(\mathbf k)$ (and likewise for $\gamma_0'$), $H = \tfrac12\sum_{\mathbf k} \big(\gamma_0(-\mathbf k)\ \gamma_0'(-\mathbf k)\big)\, h(\mathbf k)\, \big(\gamma_0(\mathbf k)\ \gamma_0'(\mathbf k)\big)^T$ with
\begin{align}
h(\mathbf k) = 2\tilde t \big[ \sin k_x\,\sigma_x - \sin k_y\,\sigma_z + (\cos k_x+\cos k_y)\sigma_y \big],
\end{align}
a gapless model with band touchings at $\mathbf k=(\pi,0),(0,\pi)$.
 
A nontrivial phase requires an on-site term coupling $\gamma_0,\gamma_0'$ on the same island, i.e.\ adding $2iM\tilde t\sum_{l,j}\gamma_0'({l,j})\gamma_0({l,j})$ to Eq.~\eqref{eq: Hinreal ap}, which is the hybridization $M$ used above. This gives
\begin{align}
h(\mathbf k) = 2\tilde t \big[ \sin k_x\,\sigma_x - \sin k_y\,\sigma_z + (\cos k_x+\cos k_y + M)\sigma_y \big],
\label{eq: bloch with mass}
\end{align}
and the Chern number
\begin{align}
C& = \frac12\Big[\mathrm{sgn}(M+2) + \mathrm{sgn}(M-2) - 2\,\mathrm{sgn}(M)\Big]
\notag\\
&= \begin{cases}
0, & M>2 \\ -1, & 0<M<2 \\ +1, & -2<M<0 \\ 0, & M<-2.
\end{cases}
\label{eq: chern number}
\end{align}
The nontrivial branch used above, $C=+1$, corresponds to $-2<M<0$, which is the bound quoted in the main text. This concludes the derivation of the effective tight-binding Hamiltonian, Eq.~\eqref{eq: Hinreal ap}, and of the bound $-2<M<0$ on $M$.
 
Now we consider the effective Hamiltonian for the coupled Majorana modes in the presence of a flux $\Phi_0$, together with the resulting Majorana zero mode.
 
We first invert the Bloch Hamiltonian of Eq.~\eqref{eq: bloch with mass}, which already includes the hybridization $M$, back to real space. Using $\gamma_0(\mathbf k) = \tfrac{1}{\sqrt N}\sum_{l,j} e^{-i(k_xla+k_yja)}\gamma_0(l,j)$ gives
\begin{widetext}
\begin{align}
&\tilde H = -i\tilde t \sum_{l,j} \Big[
2\gamma_0({l,j})\gamma_0'({l+1,j})
- \gamma_0({l,j})\gamma_0({l,j+1})
+ \gamma_0'({l,j})\gamma_0'({l,j+1})
- \gamma_0'({l,j})\gamma_0({l,j+1})
+ \gamma_0({l,j})\gamma_0'({l,j+1})
\notag\\
&- 2M\,\gamma_0'({l,j})\gamma_0({l,j})
\Big],
\label{eq: Hinreal mass}
\end{align}
\end{widetext}
i.e.\ Eq.~\eqref{eq: Hinreal ap} with the on-site hybridization term added, consistent with Eq.~\eqref{eq: bloch with mass}.
 
Threading the array with a flux $\Phi_0$ and choosing the branch-cut gauge in which every bond $\gamma_0({0,j})\gamma_0'({1,j})$ with $j\geq0$ picks up a relative sign, Eq.~\eqref{eq: Hinreal mass} becomes,
\begin{widetext}
\begin{align}
H_{\Phi_0} = &-i\tilde t \sum_{l,j} \Big[
\gamma_0'({l,j})\gamma_0'({l,j+1}) - \gamma_0({l,j})\gamma_0({l,j+1})
- \gamma_0'({l,j})\gamma_0({l,j+1})
+ \gamma_0({l,j})\gamma_0'({l,j+1})
- 2M\,\gamma_0'({l,j})\gamma_0({l,j})\Big] \notag \\
& -2i\tilde t\Big[\sum_{l\neq0,j}\gamma_0({l,j})\gamma_0'({l+1,j})
- \sum_{j\geq0}\gamma_0({0,j})\gamma_0'({1,j})
+ \sum_{j<0}\gamma_0({0,j})\gamma_0'({1,j})
\Big],
\label{eq: HPhi0 ap}
\end{align}
\end{widetext}
which is Eq.~\eqref{eq: JJ final Hamiltonian with flux} once the last three sums are combined into $\sum_{l,j} s(l,j)\,\gamma_0({l,j})\gamma_0'({l+1,j})$ with the branch-cut sign $s(l,j)$ of Eq.~\eqref{eq: branch cut}. Away from the cut this is locally gauge-equivalent to the flux-free Hamiltonian; only the enclosed flux $\Phi_0$ is physical.
 This concludes the derivation of Eq.~\eqref{eq: JJ final Hamiltonian with flux}.

\begin{figure}[t]
    \centering
    \includegraphics[width=\linewidth]{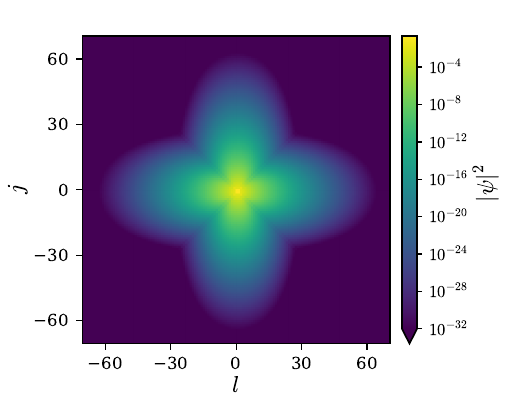}
    \caption{\justifying
    Probability density $|\psi|^2$ of the Majorana zero mode for $M=-0.7$
    on a logarithmic color scale, central part of the $201\times201$
    array.}
    \label{fig: MZM density ap}
\end{figure}

Figure~\ref{fig: MZM density ap} shows the probability density of the MZM
for $M=-0.7$, obtained by numerical diagonalization of
Eq.~\eqref{eq: JJ final Hamiltonian with flux} on a $201\times201$ array
with open boundary conditions. The density is concentrated within a few
lattice spacings of the vortex core, and its exponential tail shows a
fourfold anisotropy set by the square lattice, with a faster decay along
the diagonals, cf.\ Fig.~\ref{fig: MZM decay length}. The numerical
diagonalization yields the amplitude $\psi$ with an accuracy of
$\sim10^{-16}$, so we truncate the color scale at $|\psi|^2=10^{-32}$.

\begin{figure}[t]
    \centering
    \begin{subfigure}[b]{\linewidth}
        \centering
        \includegraphics[width=\linewidth]{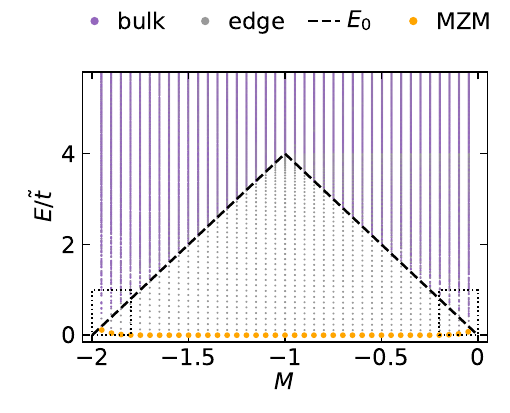}
        \caption{}
        \label{fig: spectrum full}
    \end{subfigure}
    \begin{subfigure}[b]{0.49\linewidth}
        \centering
        \includegraphics[width=\linewidth]{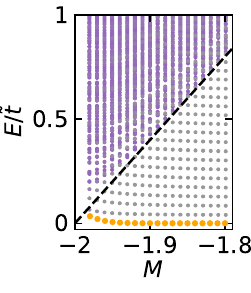}
        \caption{}
        \label{fig: spectrum zoom m2}
    \end{subfigure}
    \hfill
    \begin{subfigure}[b]{0.49\linewidth}
        \centering
        \includegraphics[width=\linewidth]{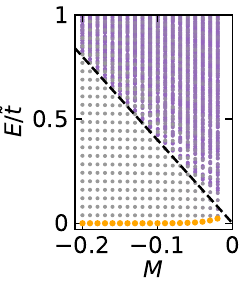}
        \caption{}
        \label{fig: spectrum zoom 0}
    \end{subfigure}
    \caption{\justifying
Low-energy spectrum of Eq.~\eqref{eq: JJ final Hamiltonian with flux} as
a function of $M$: (a) $50\times50$ array; (b), (c) zooms near $M=-2$ and
$M=0$ (dotted boxes in (a)) on a $150\times150$ array. The MZM is shown in
orange; all other levels are colored by $\rho$ (see text), from gray
($\rho\to0$, edge) to purple ($\rho\approx1$, bulk). Dashed line:
$E_0=4\tilde t\min(|M|,|M+2|)$. Apart from the MZM, all states below $E_0$
are localized at the outer boundary. Edge modes also appear above $E_0$:
the chiral edge band extends up to $4\tilde t$ for $M>-1$ and up to
$4\tilde t\sqrt{|M|(M+2)}$ for $M<-1$, and these modes do not mix with
bulk states of the same energy, which carry a different momentum along the
boundary.}
    \label{fig: spectrum ap}
\end{figure}

Figure~\ref{fig: spectrum ap}\subref{fig: spectrum full} shows the full
low-energy spectrum of Eq.~\eqref{eq: JJ final Hamiltonian with flux} as
a function of $M$, obtained by full diagonalization on a $50\times50$
array with open boundary conditions;
panels~\subref{fig: spectrum zoom m2} and~\subref{fig: spectrum zoom 0}
zoom into the regime $\xi_F\gg\xi$ near the ends of the topological
window, using a $150\times150$ array. At each $M$ the lowest level is the
MZM. Every other level is characterized by $\rho$, its weight
$\sum|\psi(l,j)|^2$ in the central square of half the linear size of the
array divided by the area fraction of this square: $\rho\approx1$ for bulk
states spread over the whole array, and $\rho\to0$ for edge modes confined
to the outer boundary. None of these levels has appreciable weight within
a distance $r<4$ of the vortex core, i.e., no other state is bound to the
vortex. Above
$E_0=4\tilde t\min(|M|,|M+2|)$ the bulk states form a continuum. Apart
from the MZM, all states below $E_0$ are chiral modes of the outer
boundary, with a level spacing set by its finite perimeter; none of them
is localized at the vortex. The MZM is therefore the only vortex-bound
state below the bulk gap. Close to $M=0$ and $M=-2$ the decay length of
the MZM, $1/|M|$ and $1/|M+2|$, becomes comparable to the array size, so
the MZM hybridizes with its partner at the outer boundary.

\section{Derivation of effective field theory}
\label{ap: Zeffective}

In this Appendix we provide the full derivation of the effective
partition sum, with the standard sine-Gordon field $\theta$ decoupling
the logarithmic vortex-vortex interaction and an additional scalar
field $\Psi$ decoupling the short-range Majorana-mediated interaction.
We thereby arrive at Eq.~\eqref{eq: sine-Gordon action} of the main text.

We start from the grand-canonical lattice-gas representation introduced
in Eq.~\eqref{eq: ZG lattice gas}.
We remark that, since configurations in the lattice-gas
representation are already unlabeled, no additional combinatorial
factor is required.

We now decouple the two interaction terms independently using the
Hubbard--Stratonovich fields $\theta(\mathbf x)$ and $\Psi(\mathbf x)$.
The field $\theta$ is dual to the logarithmic vortex-vortex interaction,
whereas $\Psi$ is dual to the Majorana-mediated interaction
$V^{\rm ferm}$ defined in Eq.~\eqref{eq: Vferm def}.

\subsection{Decoupling the logarithmic interaction}

The logarithmic Coulomb interaction can be decoupled by the standard
Hubbard--Stratonovich transformation,
\begin{align}
&\exp\Bigg[
\pi K\sum_{\mathbf r\neq\mathbf r'}
q_{\mathbf r}q_{\mathbf r'}
\ln\frac{|\mathbf r-\mathbf r'|}{\xi}
\Bigg]
\notag\\
&\qquad\sim
\int\mathcal D\theta\,
\exp\Bigg[
-\frac{1}{8\pi^2K}
\int d^2x\,(\vec\nabla\theta)^2
+i\sum_{\mathbf r}
q_{\mathbf r}\theta(\mathbf r)
\Bigg].
\label{eq: theta decoupling}
\end{align}

The field $\theta$ therefore couples linearly to the local vortex
charge $q_{\mathbf r}$.

\subsection{Decoupling the Majorana-mediated interaction}

The Majorana-mediated interaction depends only on the vortex
occupation $|q_{\mathbf r}|$ and therefore does not distinguish
vortices from antivortices. Introducing an independent scalar
Hubbard--Stratonovich field $\Psi(\mathbf x)$ gives
\begin{align}
&\exp\Bigg[
\frac{1}{2T}\sum_{\mathbf r\neq\mathbf r'}
|q_{\mathbf r}|\,|q_{\mathbf r'}|\,
V^{\rm ferm}(\mathbf r-\mathbf r')
\Bigg]\sim
\notag\\
&
\int\mathcal D\Psi\,
\exp\Bigg[
-\frac{T}{2}
\int d^2x\,d^2y\,
\Psi(\mathbf x)
[V^{\rm ferm}]^{-1}(\mathbf x-\mathbf y)
\Psi(\mathbf y)
\notag\\
&
+\sum_{\mathbf r}|q_{\mathbf r}|\Psi(\mathbf r)
\Bigg],
\label{eq: psi decoupling}
\end{align}
where $[V^{\rm ferm}]^{-1}$ denotes the 
operator 
inverse of $V^{\rm ferm}$. Note that $\Psi$ couples to vortex occupation $|q_{\mathbf r}|$ analogously to $\mu/T$. The diagonal term is absorbed into $\mu$.

\subsection{Summing over vortex charges}

Combining the two Hubbard--Stratonovich transformations with the
fugacity term, the 
sum over $q_{\mathbf r}=0,\pm1$ therefore factorizes independently
on each lattice cell:
\begin{align}
&\sum_{\{q_{\mathbf r}=0,\pm1\}}
\exp\Bigg[
i\sum_{\mathbf r}q_{\mathbf r}\theta(\mathbf r)
+
\sum_{\mathbf r}|q_{\mathbf r}|
\left(
\Psi(\mathbf r)+\frac{\mu}{T}
\right)
\Bigg]
\notag\\
&\qquad=
\prod_{\mathbf r}
\Bigg[
1+
2e^{\Psi(\mathbf r)+\mu/T}
\cos\theta(\mathbf r)
\Bigg].
\label{eq: q sum}
\end{align}
The factor of $2$ arises from the two charge states
$q_{\mathbf r}=\pm1$. Identifying $\mathcal{Y}\equiv2e^{\mu/T}$, consistent with the
definition used in the main text, the local factor in
Eq.~\eqref{eq: q sum} can be written as
$1+\mathcal{Y}\,e^{\Psi(\mathbf r)}\cos\theta(\mathbf r)$.

Consequently, after summing over the vortex charges,
the partition sum becomes
\begin{align}
&\mathcal{Z}_G
=
\int\mathcal D\theta\,\mathcal D\Psi\,
\exp\Bigg[
-\frac{1}{8\pi^2K}
\int d^2x\,(\vec\nabla\theta)^2
\notag\\
&
-\frac{T}{2}
\int d^2x\,d^2y\,
\Psi(\mathbf x)
[V^{\rm ferm}]^{-1}(\mathbf x-\mathbf y)
\Psi(\mathbf y)
\Bigg]
\notag\\
&\times
\prod_{\mathbf r}
\Big[
1+
\mathcal{Y}\,
\cos\theta(\mathbf r)\,
e^{\Psi(\mathbf r)}
\Big].
\label{eq: appendix discrete effective}
\end{align}
As mentioned briefly in the main text, this equation  defines the regularization of Eq.~\eqref{eq: sine-Gordon action} of the main text, in particular when the factor $1+
\mathcal{Y}\,
\cos\theta(\mathbf r)\,
e^{\Psi(\mathbf r)}$ is negative.

\subsection{Continuum limit and final action}

There are
$\mathcal A/\xi^2$ lattice cells, each of area $\xi^2$. In the
continuum limit, the lattice sum is replaced by
\begin{equation}
\sum_{\mathbf r}
\longrightarrow
\frac{1}{\xi^2}\int d^2x,
\end{equation}
while the product over lattice cells is converted into an exponential,
\begin{align}
&\prod_{\mathbf r}
\Big[
1+
\mathcal{Y}\,
\cos\theta(\mathbf r)\,
e^{\Psi(\mathbf r)}
\Big]
=
\exp\Bigg\{
\sum_{\mathbf r}
\ln\Big[
1+
\mathcal{Y}\,
\cos\theta(\mathbf r)\,
e^{\Psi(\mathbf r)}
\Big]
\Bigg\}
\notag\\
&\longrightarrow
\exp\Bigg\{
\frac{1}{\xi^2}
\int d^2x\,
\ln\Big[
1+
\mathcal{Y}\,
\cos\theta(\mathbf x)\,
e^{\Psi(\mathbf x)}
\Big]
\Bigg\}.
\end{align}

We therefore obtain
\begin{equation}
\mathcal{Z}_G
\equiv
\int\mathcal D\theta\,\mathcal D\Psi\,
e^{-S[\theta,\Psi]},
\end{equation}
with
\begin{align}
S[\theta,\Psi]
&=
\int d^2x
\Bigg[
\frac{1}{8\pi^2K}
(\vec\nabla\theta)^2
\notag\\
&\qquad
+\frac{T}{2}
\int d^2y\,
\Psi(\mathbf x)
[V^{\rm ferm}]^{-1}(\mathbf x-\mathbf y)
\Psi(\mathbf y)
\notag\\
&\qquad
-\frac{1}{\xi^2}
\ln\Big(
1+
\mathcal{Y}
\cos\theta(\mathbf x)
e^{\Psi(\mathbf x)}
\Big)
\Bigg].
\label{eq: appendix final action}
\end{align}

This is precisely the effective sine-Gordon-type action quoted in
Eq.~\eqref{eq: sine-Gordon action} of the main text, with  fugacity $\mathcal{Y}=2e^{\mu/T}$ 
and $[V^{\rm ferm}]^{-1}$ defined by
Eq.~\eqref{eq: Vinv def}.

\subsection{Regime of two minima}
\label{app:TwoMin}

Here we analyze the bare potential at $\theta=0$, Eq.~\eqref{eq:VthetaPsi},
\begin{equation}
V(0,\Psi)=b\Psi^2-\ln\big(1+\mathcal Y e^{\Psi}\big),
\end{equation}
and derive the window of two minima, Eq.~\eqref{eq:WindowFor2Min}, the
stationary points, and the first-order condition, Eq.~\eqref{eq:FirstOrder}.
The same expressions apply to the potential of the instanton analysis,
Eq.~\eqref{eq: instanton effective action}, which is the bare potential.

\paragraph{Window of two minima.}
At the boundary of the range in $b$ or $\mathcal Y$ for which $V(0,\Psi)$
displays two minima, $\partial_\Psi V(0,\Psi)$ has an extremum at zero,
\begin{subequations}
\begin{align}
    0 & = \partial_\Psi V(0,\Psi)  = 2 b\Psi - s,\label{eq:dV}\\
    0 & = \partial_\Psi^2 V(0,\Psi)  = 2 b - s (1-s), \label{eq:d2V}
\end{align}
\end{subequations}
where $s = \mathcal Y e^{\Psi}/(1+\mathcal Y e^{\Psi})$. Hence
\begin{equation}
    \Psi  \stackrel{\eqref{eq:dV}}{=} \frac{s }{2b} \stackrel{\eqref{eq:d2V}}{=} \frac{1}{1-s},
\end{equation}
and, using the definition of $s$,
\begin{equation}
    \mathcal Y = (\Psi-1)e^{-\Psi},\qquad
    b = \frac{\Psi-1}{2 \Psi^2}.
\end{equation}
Real solutions of the second equation exist for $b\leq 1/8$,
\begin{equation}
    \Psi_\pm = \frac{1 \pm \sqrt{1 - 8 b}}{4b} \simeq \begin{cases}
        \frac{1}{2b}-1, & +, \\
        1 + 2b, & -,
    \end{cases}
\end{equation}
where at the $\simeq$ sign we expanded in small $b$, the main regime of
interest. The window in $\mathcal Y$ at fixed $b$ is thus bounded by
\begin{equation}
     \mathcal Y  = (\Psi_\pm-1)e^{-\Psi_\pm} \simeq \begin{cases}
         \frac{e}{2b} e^{-\frac{1}{2b}}, &+, \\
         2b/e, &-,
     \end{cases}
\end{equation}
and, by inversion, the window in $b$ at fixed small $\mathcal Y$ is
\begin{equation}
    b_< \simeq \frac{e\mathcal Y}{2}, \qquad
    b_> \simeq \frac{1/2}{\ln(1/\mathcal Y)+\ln\ln(1/\mathcal Y) + 1},
\end{equation}
which to leading order is Eq.~\eqref{eq:WindowFor2Min}.

\paragraph{Stationary points.}
Inside the window, the extremum condition $2b\Psi=s$ has three roots
$\Psi_1<\Psi_0<\Psi_2$. Expanding in the appropriate limit for each root
(small $\Psi_1$; $\mathcal Y e^{\Psi_2}\gg1$; one Newton iteration for the
barrier top $\Psi_0$) gives
\begin{align}
\Psi_1&\approx\frac{\mathcal Y}{2b}+\frac{1-2b}{4b^2}\mathcal Y^2, \\
\Psi_2&\approx\frac1{2b}-\frac{e^{-1/2b}}{2b\,\mathcal Y}, \\
\Psi_0&\approx\ln\frac1{\mathcal Y}+\ln\frac{2b\ln(1/\mathcal Y)}{1-2b\ln(1/\mathcal Y)} ,
\end{align}
valid for $b_<\ll b<b_>$.

\paragraph{Splitting, barrier, and first-order transition.}
Evaluating $V(0,\Psi)$ at the stationary points gives the splitting between
the two minima and, near degeneracy, the barrier height,
\begin{align}
\xi^2\epsilon=V(0,\Psi_2)-V(0,\Psi_1)&\approx\ln\frac1{\mathcal Y}+\mathcal Y-\frac1{4b}-\frac{e^{-1/2b}}{\mathcal Y},\\
\xi^2\Delta V=V(0,\Psi_0)-V(0,\Psi_2)&\approx\frac1{16b}-\ln2 .
\end{align}
The last term of $\xi^2\epsilon$ stems from
$\ln(1+\mathcal Ye^{\Psi_2})\simeq\ln\mathcal Y+\Psi_2+e^{-\Psi_2}/\mathcal Y$.
Near degeneracy $e^{-1/2b}/\mathcal Y\simeq\mathcal Y$, so it cancels the term
$\mathcal Y$ from $V(0,\Psi_1)\simeq-\mathcal Y$, and
\begin{equation}
\xi^2\epsilon\simeq\ln\frac1{\mathcal Y}-\frac1{4b}.
\end{equation}
The two minima are degenerate, $\epsilon=0$, at
$b=\tfrac14\big/\ln(1/\mathcal Y)$, which is
Eq.~\eqref{eq:FirstOrder}, i.e.\ $2b\ln(1/\mathcal Y)\simeq\tfrac12$ and
$b\simeq b_>/2$, well inside the window of two minima.

\paragraph{Turning point.}
The turning point $\Psi_\star$ on the $\Psi_1$ side of the barrier, defined by
$V(0,\Psi_\star)=V(0,\Psi_2)$ and needed for the bounce, follows from
$\ln(1+\mathcal Y e^{\Psi})\approx0$ for $\Psi<\Psi_0$,
\begin{equation}
\Psi_\star\approx\sqrt{\frac1b\ln\frac1{\mathcal Y}-\frac1{4b^2}} .
\end{equation}

\begin{figure*}
    \centering
    \begin{minipage}[t]{0.32\textwidth}
        \raggedright (a)\\
        \includegraphics[width=\linewidth]{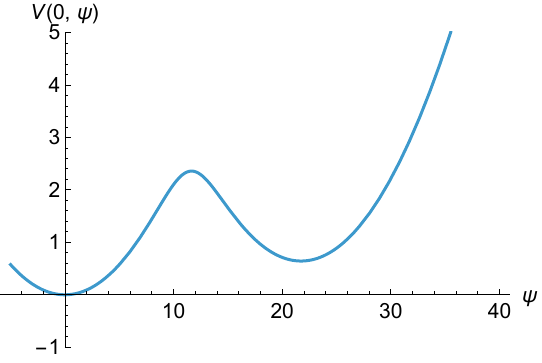}
    \end{minipage}\hfill
    \begin{minipage}[t]{0.32\textwidth}
        \raggedright (b)\\
        \includegraphics[width=\linewidth]{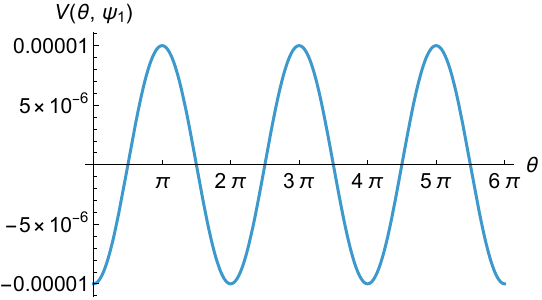}
    \end{minipage}\hfill
    \begin{minipage}[t]{0.32\textwidth}
        \raggedright (c)\\
        \includegraphics[width=\linewidth]{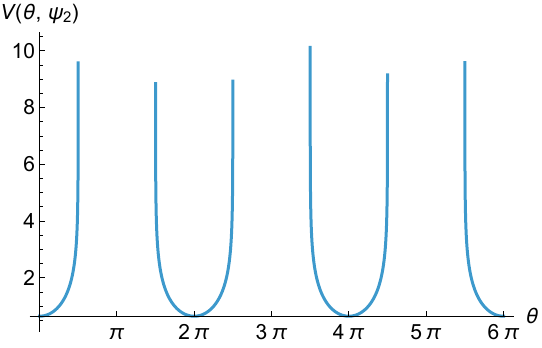}
    \end{minipage}
    \caption{\justifying
    Cuts of the potential illustrating its stationary-point structure.
    (a) $V(0,\Psi)$ along the $\Psi$ direction, showing the minima
    $\Psi_1$, $\Psi_2$ and the barrier top $\Psi_0$.
    (b) The $\theta$ dependence of the potential at $\Psi=\Psi_1$.
    (c) The $\theta$ dependence at $\Psi=\Psi_2$.
    The plots are shown for $\mathcal Y=10^{-5}$ and $b=0.023$, slightly above the first-order value $b\simeq0.0217$, Eq.~\eqref{eq:FirstOrder}, so that $\Psi_1$ is the true and $\Psi_2$ the metastable minimum.}
    \label{fig: potential cuts}
\end{figure*}

The structure of the potential is illustrated in
Fig.~\ref{fig: potential cuts}: panel~(a) shows $V(0,\Psi)$ with the three
stationary points $\Psi_1<\Psi_0<\Psi_2$; panels~(b) and (c) show the
$\theta$ dependence at $\Psi_1$ and $\Psi_2$.

\section{Renormalization group analysis}
\label{ap: RG}

Here we derive the one-loop RG equations for the action of
Eq.~\eqref{eq: full symmetry action} by the standard momentum-shell
procedure, and show how they reduce to the single-fugacity system of
Eq.~\eqref{eq: reduced RG system} quoted in the main text.

\subsection{The general symmetry-allowed action}
Symmetry alone --- periodicity of $\theta$ under $\theta\to\theta+2\pi$
and the structure of the coupling of the fluctuating chemical potential
to the vortex fugacity --- permits, beyond the single cosine term of
Eq.~\eqref{eq: sine-Gordon action}, a full tower of harmonics
$\mathcal Y_{\alpha,n}\cos(n\theta)e^{\alpha\Psi}$, together with
gradient-coupled operators $\gamma_{\alpha,n}(\nabla\theta)^2$ and
$\eta_{\alpha,n}(\nabla\Psi)^2$ dressed by the same harmonics. These
operators are not forbidden by any symmetry of the problem and are
generically generated under RG even if absent in the bare action, so we
retain them from the outset; we take $O(\gamma),O(\eta)\le O(\mathcal Y^2)$
throughout, since operators of this type are generated only at second
order in $\mathcal Y$ (confirmed self-consistently in
Eq.~\eqref{eq: full RG system}). The most general local action
consistent with these symmetries is
\begin{align}
 & S[\theta,\Psi]=S_{\rm C}[\theta]+S_{\rm SR}[\Psi]+S_{\rm int}[\theta,\Psi],\notag\\
 & S_{\rm int}[\theta,\Psi]=\int d^{2}x\sum_{\alpha,n}\Bigg(-\frac{\lambda^{2}}{\xi^{2}}\ln\Big[1+\mathcal{Y}_{\alpha,n}e^{\alpha\Psi}\cos(n\theta)\Big]\notag\\
&+e^{\alpha\Psi}\cos(n\theta)\Big(\gamma_{\alpha,n}(\nabla\theta)^{2}+\eta_{\alpha,n}(\nabla\Psi)^{2}\Big)\Bigg),
\label{eq: full symmetry action}
\end{align}
with $S_{\rm C}$ as in Eq.~\eqref{eq: sine-Gordon action}, $S_{\rm SR}$ as in
Eq.~\eqref{eq: full theta Psi action}, $\alpha,n\in\mathbb N_0$ with $(\alpha,n)\neq(0,0)$,
$\mathcal Y_{1,1}(0)\equiv\mathcal Y(0)$ and $\lambda(0)=1$.
 
\subsection{Kernel of the fluctuating chemical potential}
 
$S_{\rm SR}$ follows from Fourier transforming $V^{\rm ferm}$ of
Eq.~\eqref{eq: Vferm def}. For $x=\tfrac{E_0}{T}e^{-r/\xi_F}\gg1$ (i.e.\
$r\ll\xi_F\ln(E_0/T)$, satisfied during the first stage of the flow for
$E_0/T\gg1$), $\ln\cosh x\simeq x-\ln2$; the constant contributes only at
$q=0$, and
\begin{align}
&V^{\rm ferm}(q)=T\int d^2r\,\ln\Big(\cosh\Big[\frac{E_0}{T}e^{-r/\xi_F}\Big]\Big)e^{i\mathbf q\mathbf r}
\notag\\
&\simeq2\pi E_0\int_0^\infty dr\,r\,e^{-r/\xi_F}J_0(qr)
=\frac{2\pi E_0\,\xi_F^2}{(1+q^2\xi_F^2)^{3/2}}.
\end{align}
With $S_{\rm SR}=\tfrac T2\int\Psi\,[V^{\rm ferm}]^{-1}\Psi$ this gives
Eq.~\eqref{eq: full theta Psi action} with
$\kappa(0)=\frac{E_0\xi}{2\pi T\xi_F}$ and $\mathcal M(0)=\frac{\xi}{\xi_F}$.
\subsection{Momentum-shell renormalization}
 
One RG step consists of (a) integrating out the fast modes,
$\theta=\theta_s+\theta_f$, $\Psi=\Psi_s+\Psi_f$, with slow modes at
$0<p<\Lambda/u$ and fast modes at $\Lambda/u\le p<\Lambda$, $\Lambda=1/\xi$,
$u=e^l$,
\begin{equation}
S_{\rm eff}[\theta_s,\Psi_s]=S_{\rm C}[\theta_s]+S_{\rm SR}[\Psi_s]+\langle S_{\rm int}\rangle_f-\tfrac12\big(\langle S_{\rm int}^2\rangle_f-\langle S_{\rm int}\rangle_f^2\big),
\end{equation}
and (b) rescaling $\mathbf x\to\mathbf x\,e^{l}$ back to the cutoff $\Lambda$.
Step (b) alone gives $d\lambda/dl=\lambda$, $d\kappa/dl=\kappa$,
$d\mathcal M/dl=\mathcal M$ and leaves $\gamma,\eta$ unchanged (the two
gradients compensate the $e^{2l}$ of the measure); $\mathcal Y_{\alpha,n}$,
$\gamma_{\alpha,n}$, $\eta_{\alpha,n}$ change only in step (a). We work to
$O(\mathcal Y^2)$, expanding $\ln(1+y)=y-\tfrac12y^2+\dots$. The fast-mode
averages are controlled by the shell correlators
\begin{align}
&C_\theta(r)=2\pi K\int_{\Lambda/u}^\Lambda\frac{dp}{p}J_0(pr),\notag\\
&C_\Psi(r)=\frac{2\pi\kappa}{\xi}\int_{\Lambda/u}^\Lambda dp\,\frac{p\,J_0(pr)}{(\mathcal M^2\xi^{-2}+p^2)^{3/2}}.
\end{align}
Using $\langle e^{i n\theta_f+\alpha\Psi_f}\rangle_f=e^{-\frac{n^2}2C_\theta(0)+\frac{\alpha^2}2C_\Psi(0)}$
for single vertices and the analogous two-point contraction (at separation
$\mathbf r=\mathbf x-\mathbf y$) for the terms from $S_{\rm int}^2$, step (a) gives
\begin{widetext}
\begin{subequations}
\label{eq: primed couplings exact}
\begin{align}
\lambda^2\mathcal Y'_{\alpha,n} ={}& e^{\frac{\alpha^2}2C_\Psi(0)-\frac{n^2}2C_\theta(0)}\Big[\lambda^2\mathcal Y_{\alpha,n}-\xi^2\gamma_{\alpha,n}\langle(\nabla\theta_f)^2\rangle-\xi^2\eta_{\alpha,n}\langle(\nabla\Psi_f)^2\rangle\Big] \notag\\
&+\sum_\pm\sum_{\alpha_1,\alpha_2,n_1,n_2}\frac{\mathcal Y_{\alpha_1,n_1}\mathcal Y_{\alpha_2,n_2}}4\,\delta_{\alpha,\alpha_1+\alpha_2}\delta_{n,n_1\pm n_2}\,
e^{\frac{\alpha_1^2+\alpha_2^2}2C_\Psi(0)-\frac{n_1^2+n_2^2}2C_\theta(0)}\notag\\
&\times\Big[-\lambda^2e^{\alpha_1\alpha_2C_\Psi(0)\mp n_1n_2C_\theta(0)}-\frac{\lambda^4}{\xi^2}\int_{\mathbf r}\big(1-e^{\alpha_1\alpha_2C_\Psi(r)\mp n_1n_2C_\theta(r)}\big)\Big], \\[1ex]
\gamma'_{\alpha,n} ={}& e^{\frac{\alpha^2}2C_\Psi(0)-\frac{n^2}2C_\theta(0)}\gamma_{\alpha,n} - \sum_\pm\sum_{\alpha_1,\alpha_2,n_1,n_2}\frac{\mathcal Y_{\alpha_1,n_1}\mathcal Y_{\alpha_2,n_2}}4\,\delta_{\alpha,\alpha_1+\alpha_2}\delta_{n,n_1\pm n_2}\,
e^{\frac{\alpha_1^2+\alpha_2^2}2C_\Psi(0)-\frac{n_1^2+n_2^2}2C_\theta(0)} \notag\\
&\times\frac{\lambda^4}{\xi^4}\,\frac{(n_1\mp n_2)^2}{16}\int_{\mathbf r}\mathbf r^2\big[1-e^{\alpha_1\alpha_2C_\Psi(r)\mp n_1n_2C_\theta(r)}\big], \\[1ex]
\eta'_{\alpha,n} ={}& e^{\frac{\alpha^2}2C_\Psi(0)-\frac{n^2}2C_\theta(0)}\eta_{\alpha,n} + \sum_\pm\sum_{\alpha_1,\alpha_2,n_1,n_2}\frac{\mathcal Y_{\alpha_1,n_1}\mathcal Y_{\alpha_2,n_2}}4\,\delta_{\alpha,\alpha_1+\alpha_2}\delta_{n,n_1\pm n_2}\,
e^{\frac{\alpha_1^2+\alpha_2^2}2C_\Psi(0)-\frac{n_1^2+n_2^2}2C_\theta(0)} \notag\\
&\times\frac{\lambda^4}{\xi^4}\,\frac{(\alpha_1-\alpha_2)^2}{16}\int_{\mathbf r}\mathbf r^2\big[1-e^{\alpha_1\alpha_2C_\Psi(r)\mp n_1n_2C_\theta(r)}\big].
\end{align}
\end{subequations}
\end{widetext}
The on-site term $\propto\lambda^2$ stems from the $-\tfrac12y^2$ term of the
logarithm, the term $\propto\lambda^4$ from the second cumulant.
For a thin shell, $\int_{\Lambda/u}^\Lambda dp\,f(p)\simeq\Lambda l\,f(\Lambda)$,
and with $\mathcal M\ll1$
\begin{align}
&C_\theta(r)=2\pi K\,l\,J_0(\Lambda r),\\
&C_\Psi(r)=2\pi\kappa\,l\,J_0(\Lambda r),\\
&\xi^2\langle(\nabla\theta_f)^2\rangle=2\pi K\,l,\\
&\xi^2\langle(\nabla\Psi_f)^2\rangle=2\pi\kappa\,l,
\end{align}
while the position integrals reduce to
$\int_{\mathbf r}\big(1-e^{\cdots}\big)\simeq-\big(\alpha_1\alpha_2\,2\pi\kappa\mp n_1n_2\,2\pi K\big)\,l\,\xi^2C_1$
and
$\int_{\mathbf r}\mathbf r^2\big(1-e^{\cdots}\big)\simeq-\big(\alpha_1\alpha_2\,2\pi\kappa\mp n_1n_2\,2\pi K\big)\,l\,\xi^4C_2$,
with $C_1=\int d^2x\,J_0(x)$, $C_2=\int d^2x\,x^2J_0(x)$. Differentiating
Eq.~\eqref{eq: primed couplings exact} at $l=0$ and adding step (b) gives
the one-loop system
\begin{widetext}
\begin{align}
\begin{cases}
\dfrac{d\mathcal Y_{\alpha,n}}{dl} = \big(\alpha^2\pi\kappa-n^2\pi K\big)\mathcal Y_{\alpha,n} - \dfrac{2}{\lambda^2}\big(\pi K\,\gamma_{\alpha,n}+\pi\kappa\,\eta_{\alpha,n}\big) + \mathcal C_1\displaystyle\sum_\pm\sum_{\alpha_1,\alpha_2,n_1,n_2}\mathcal Y_{\alpha_1,n_1}\mathcal Y_{\alpha_2,n_2}\,\delta_{\alpha,\alpha_1+\alpha_2}\delta_{n,n_1\pm n_2}\big(\alpha_1\alpha_2\pi\kappa\mp n_1n_2\pi K\big), \\[1.5ex]
\dfrac{d\gamma_{\alpha,n}}{dl} = \big(\alpha^2\pi\kappa-n^2\pi K\big)\gamma_{\alpha,n} + \mathcal C_2\lambda^4\displaystyle\sum_\pm\sum_{\alpha_1,\alpha_2,n_1,n_2}\mathcal Y_{\alpha_1,n_1}\mathcal Y_{\alpha_2,n_2}\,\delta_{\alpha,\alpha_1+\alpha_2}\delta_{n,|n_1\pm n_2|}(n_1\mp n_2)^2\big(\alpha_1\alpha_2\pi\kappa\mp n_1n_2\pi K\big), \\[1.5ex]
\dfrac{d\eta_{\alpha,n}}{dl} = \big(\alpha^2\pi\kappa-n^2\pi K\big)\eta_{\alpha,n} - \mathcal C_2\lambda^4\displaystyle\sum_\pm\sum_{\alpha_1,\alpha_2,n_1,n_2}\mathcal Y_{\alpha_1,n_1}\mathcal Y_{\alpha_2,n_2}\,\delta_{\alpha,\alpha_1+\alpha_2}\delta_{n,|n_1\pm n_2|}(\alpha_1-\alpha_2)^2\big(\alpha_1\alpha_2\pi\kappa\mp n_1n_2\pi K\big), \\[1.5ex]
\dfrac{d\lambda}{dl} = \lambda, \qquad \dfrac{d\kappa}{dl} = \kappa, \qquad \dfrac{d\mathcal M}{dl} = \mathcal M, \qquad \dfrac{dK}{dl} = 0,
\end{cases}
\label{eq: full RG system}
\end{align}
\end{widetext}
with $\mathcal C_1=\tfrac12(\lambda^2C_1-1)$ and $\mathcal C_2=\tfrac1{32}C_2$.
After bringing the quadratic terms to a common normalization, the
$O(\mathcal Y^2)$ term in the $\mathcal Y_{\alpha,n}$ equation is subleading
to the corresponding terms in the $\gamma,\eta$ equations and can be dropped.
 
Three structural features of Eq.~\eqref{eq: full RG system} control which
couplings matter. First, the Kronecker deltas only connect couplings with
$\alpha+n$ even, so with the single seed $\mathcal Y_{1,1}$ no odd-parity
coupling is ever generated. Second, $\gamma_{\alpha,n},\eta_{\alpha,n}$ are
first generated (at $O(\mathcal Y^2)$) only at $(\alpha,n)=(2,0),(2,2)$ and
feed back into the $\mathcal Y_{\alpha,n}$ equation only at the same order,
so they remain subdominant to the leading fugacity. Third, couplings with
$\alpha=0$ do not involve $\kappa$; since $\mathcal M/\kappa=2\pi T/E_0\ll1$ is
not renormalized while $\mathcal M\kappa\propto e^{2l}$ grows, these terms are
either RG irrelevant or, if absent initially, never generated.
 
As a result, at leading order the flow is controlled by the single fugacity
$\mathcal Y_{1,1}$ together with $\kappa$, $\mathcal M$, $\lambda$,
and Eq.~\eqref{eq: full RG system} truncates to
Eq.~\eqref{eq: reduced RG system} of the main text,
$d(\lambda^2 \mathcal Y_{1,1})/dl=(2+\pi\kappa-\pi K)\lambda^2\mathcal Y_{1,1}$,
$d\kappa/dl=\kappa$, $d\mathcal M/dl=\mathcal M$, $dK/dl=0$,
where the fugacity of the main text is $\mathcal Y=\lambda^2\mathcal Y_{1,1}$
[with $\mathcal Y(0)=\mathcal Y_{1,1}(0)$ since $\lambda(0)=1$]. 

\section{Instanton calculations}
\label{ap: instanton}
In this appendix we present the full calculation of the decay rate $\Gamma$, Eq.~\eqref{eq: decay rate result} of the main text. Throughout, $V(\theta,\Psi)$ is the dimensionless bare potential, Eq.~\eqref{eq:VthetaPsi}, so that the potential density is $V/\xi^2$.

\subsection{Splitting, barrier and wall tension}
\label{ap: wall tension}
Using $\ln(1+\mathcal Ye^\Psi)=\frac12\big(\Psi-\ln\frac1{\mathcal Y}\big)+\ln2+\ln\cosh\frac12\big(\Psi-\ln\frac1{\mathcal Y}\big)$, the bare potential at $\theta=0$ reads
\begin{align}
V(0,\Psi)={}&b\Big(\Psi-\ln\frac1{\mathcal Y}\Big)^2-\ln\cosh\frac12\Big(\Psi-\ln\frac1{\mathcal Y}\Big)\notag\\
&+2b\Big(\ln\frac1{\mathcal Y}-\frac1{4b}\Big)\Big(\Psi-\ln\frac1{\mathcal Y}\Big)+{\rm const}.
\end{align}
The first line is even in $\Psi-\ln(1/\mathcal Y)$, the second is a linear tilt. The two minima are therefore degenerate if and only if the tilt vanishes, i.e.\ exactly on the first-order line $4b\ln(1/\mathcal Y)=1$, where $V(0,\Psi)$ is symmetric about the barrier $\Psi_0=\ln(1/\mathcal Y)$. Up to corrections of relative order $\mathcal Y$, its minima are $\Psi_1\simeq0$ and $\Psi_2\simeq2\ln(1/\mathcal Y)=1/(2b)$, $\Psi_\star=\Psi_1$, and
\begin{align}
&V(0,\Psi)-V(0,\Psi_1)=\notag\\
&b\Big(\ln\frac1{\mathcal Y}-\Big|\Psi-\ln\frac1{\mathcal Y}\Big|\Big)^2-\ln\Big(1+e^{-|\Psi-\ln(1/\mathcal Y)|}\Big),
\label{eq: V on line}
\end{align}
i.e.\ two parabolic wells meeting at the barrier, with the cusp smoothed by the last term, which is exponentially localized at $|\Psi-\Psi_0|\lesssim1$. To first order in the tilt, the splitting is the tilt times $\Psi_2-\Psi_1$, and the barrier follows from Eq.~\eqref{eq: V on line},
\begin{equation}
\xi^2\epsilon\simeq\ln\frac1{\mathcal Y}-\frac1{4b},\qquad
\xi^2\Delta V\simeq\frac1{16b}-\ln2\simeq\frac14\ln\frac1{\mathcal Y},
\end{equation}
which is Eq.~\eqref{eq: eps DeltaV} of the main text. For $\epsilon\ll\Delta V$ the stationary points shift only by a relative amount $\mathcal O(\epsilon/\Delta V)$, so to zeroth order in $\epsilon$ the wall tension is evaluated on the first-order line. There, Eq.~\eqref{eq: V on line} gives
\begin{align}
&\int_{\Psi_1}^{\Psi_2}d\Psi\sqrt{V(0,\Psi)-V(0,\Psi_1)}
=\notag\\
&\sqrt b\,\ln^2\frac1{\mathcal Y}\Big[1-\frac{\pi^2}{3\ln^2(1/\mathcal Y)}+\mathcal O\Big(\frac1{\ln^3(1/\mathcal Y)}\Big)\Big].
\end{align}
The leading term is the contribution of the two parabolic wells, $2\sqrt b\int_0^{\Psi_0}d\Psi\,(\Psi_0-|\Psi-\Psi_0|)$; the correction comes from the smoothing of the cusp. With $b=\mathcal M^3/(8\pi^2\kappa)$ this gives Eqs.~\eqref{eq: wall tension}--\eqref{eq: bounce radius}.

\subsection{Separation of scales}
\label{ap: scales}
The bounce involves three lengths, see Fig.~\ref{fig: coleman-bounce}.

\emph{Wall thickness.} Across the wall $\Psi$ changes from $\Psi_1$ to $\Psi_2$. For a flat wall at $\theta=0$ and $\epsilon=0$, the first integral of the Euler--Lagrange equation of Eq.~\eqref{eq: instanton effective action} reads
\begin{equation}
\frac12\,\frac{3\mathcal M}{8\pi^2\kappa}\Big(\frac{d\Psi}{dx}\Big)^2=\frac{V(0,\Psi)-V(0,\Psi_2)}{\xi^2}.
\end{equation}
This is the energy conservation $\frac12m\dot x^2=V-E$ of a particle of mass $m=\frac{3\mathcal M}{8\pi^2\kappa}$ with coordinate $\Psi$ and time $x$. Near the minima, where $V(0,\Psi)-V(0,\Psi_{1,2})\simeq b(\Psi-\Psi_{1,2})^2$, the particle approaches the vacua exponentially, $|\Psi-\Psi_{1,2}|\propto e^{-|x|/(\sqrt{3/2}\,\xi_F)}$. The travel time from $\Psi_1$ to $\Psi_2$ therefore diverges logarithmically, and the wall thickness is defined by stopping at a distance $\delta$ from the vacua. From Eq.~\eqref{eq: V on line},
\begin{align}
\ell_{\rm w}(\delta)&=\xi\sqrt{\frac{3\mathcal M}{16\pi^2\kappa}}\int_{\Psi_1+\delta}^{\Psi_2-\delta}\frac{d\Psi}{\sqrt{V(0,\Psi)-V(0,\Psi_2)}}\notag\\
&=\sqrt6\,\xi_F\Big[\ln\frac{\ln(1/\mathcal Y)}{\delta}+\mathcal O\Big(\frac1{\ln^2(1/\mathcal Y)}\Big)\Big].
\label{eq: wall thickness}
\end{align}
The logarithm only counts the decay lengths $\sqrt{3/2}\,\xi_F$ needed to cover the large jump $\Psi_2-\Psi_1\simeq2\ln(1/\mathcal Y)$ to accuracy $\delta$. The energy density of the wall, $2[V(0,\Psi)-V(0,\Psi_2)]/\xi^2$, decays as $e^{-2|x|/(\sqrt{3/2}\,\xi_F)}$ away from the wall center, so the wall tension is accumulated within a distance of order $\xi_F$. The physical width of the wall is therefore set by the prefactor $\sqrt6\,\xi_F$, which we use below.

\emph{Bubble radius.} $R_B=\sigma_\Psi/\epsilon$, Eq.~\eqref{eq: bounce radius}. With Eq.~\eqref{eq: eps DeltaV}, the wall tension, Eq.~\eqref{eq: wall tension}, is $\sigma_\Psi\simeq2\Delta V\sqrt{3/2}\,\xi_F$, so
\begin{equation}
\frac{R_B}{\sqrt6\,\xi_F}=\frac{\Delta V}{\epsilon}.
\label{eq: RB over wall}
\end{equation}

\emph{Kink width.} Inside the bubble, $\Psi\simeq\Psi_1\simeq0$ and the potential for $\theta$ is $-\mathcal Y\cos\theta/\xi^2$. A kink in $\theta$ has width $R_k = \xi/(2\pi\sqrt{K\mathcal{Y}})$.

Using the bare values of $\mathcal M$ and $\kappa$ and $\ln(1/\mathcal Y)=\pi\frac{E_0}{T}(\xi_F/\xi)^2$, the remaining ratios are
\begin{align}
&\frac{R_k}{\sqrt6\,\xi_F}
=\frac{1}{2\pi\sqrt{6K}}\,\frac{\xi}{\xi_F}\,
e^{\frac{\pi}{2}\frac{E_0}{T}(\xi_F/\xi)^2},
\label{eq: kink over wall}\\
\frac{R_B}{R_k}
&=\frac{\sqrt{6K}\,\mathcal M^2}{4\pi\kappa}\,\frac{\sqrt{\mathcal Y}\,\ln^2(1/\mathcal Y)}{\epsilon\xi^2}\notag\\
&=\sqrt{\frac{3K}{2}}\,\pi^2\,\frac{E_0}{T}\Big(\frac{\xi_F}{\xi}\Big)^3
\frac{e^{-\frac{\pi}{2}\frac{E_0}{T}(\xi_F/\xi)^2}}{\epsilon\xi^2}.
\label{eq: RB over kink}
\end{align}

Eq.~\eqref{eq: RB over wall} shows that $R_B\gg\sqrt6\,\xi_F$ is precisely the thin-wall condition $\epsilon\ll\Delta V$.

Eq.~\eqref{eq: kink over wall} is exponentially large for any $\epsilon$. Hence $\theta$ cannot follow $\Psi$ across the wall. This justifies constructing the bounce at fixed $\theta=0$ and treating the wall as a Dirichlet boundary for $\theta$.

From Eq.~\eqref{eq: RB over kink} we conclude that for exponentially small $\epsilon$, where $R_B \gg R_k$, kinks fit into the bubble, and $\theta$ remains a free field because the cosine term is irrelevant for $K>2/\pi$, as discussed in the main text.

\subsection{Contribution from superfluid inside the bubble}
The free $\theta$ field inside the bubble contributes to the single-bounce amplitude through its partition function on a disk of radius $R_B$ with Dirichlet boundary conditions,
\begin{equation}
Z_\theta^{\rm free}(R_B)=\int_{\theta|_{r=R_B}=0}\mathcal D\theta\,e^{-S^{(2)}[\theta]}=\Big[\det\Big(-\frac{\nabla^2}{4\pi^2K}\Big)\Big]^{-1/2}.
\end{equation}
In units of $\xi$, the eigenfunctions are $\theta_{mn}(r,\phi)=J_m(j_{m,n}r/r_B)e^{im\phi}$, with $j_{m,n}$ the zeros of $J_m$, $r_B=R_B/\xi$, and eigenvalues $j_{m,n}^2/r_B^2$, so that
\begin{equation}
\det\Big(-\frac{\nabla^2}{4\pi^2K}\Big)=\prod_{m,n}\Big(\frac{1}{4\pi^2K}\frac{j_{m,n}^2}{r_B^2}\Big).
\end{equation}
{This functional determinant is an infinite product over the modes of $\theta$ in the disk and requires regularization. With an explicit short-distance cutoff $\xi$, its logarithm contains terms proportional to the area and to the perimeter of the bubble, with non-universal coefficients that depend on the details of the cutoff procedure, and a logarithmic term with a universal coefficient. The area and perimeter terms have the same dependence on $R_B$ as the bulk and wall contributions to the bounce action. They therefore merely renormalize $\epsilon$ and $\sigma_\Psi$ and do not affect the physics; we understand $\epsilon$ and $\sigma_\Psi$ to include these corrections. The only contribution that cannot be absorbed in this way is the universal logarithm. We extract it using $\zeta$-function regularization, which discards the non-universal terms automatically,} and
using $Z_D(s)=\sum_{m,n}j_{m,n}^{-2s}$ with $Z_D(0)=1/6$ for the Dirichlet disk, this gives Eq.~\eqref{eq: Z theta free} of the main text.  Since it depends on $R_B$, it also enters the integral over the unstable radial mode. Writing $r=r_B+\Delta r$, with $B(r_B+\Delta r)-B(r_B)=-\pi\epsilon\xi^2(\Delta r)^2$, and expanding $(r_B+\Delta r)^{1/6}$ to second order in $\Delta r$, the Gaussian integral over $\Delta r$, continued in the standard way, gives
\begin{align}
&\int d(\Delta r)\,e^{\pi\epsilon\xi^2(\Delta r)^2}(r_B+\Delta r)^{1/6}
=\notag \\ 
&\frac{i}{\sqrt{\epsilon\xi^2}}\,r_B^{1/6}\Big[1+\frac{5}{144\pi\epsilon\xi^2 r_B^2}\Big],
\end{align}
where the correction is negligible, since $\pi\epsilon\xi^2r_B^2=B\gg1$.

\subsection{Fluctuation determinant about the bounce}
Expanding $\Psi=\Psi_B+\eta$ about the bounce profile $\Psi_B(r)$ to quadratic order gives $S[\eta]=\tfrac12\int d^2x\,\eta\,\mathcal H_B\,\eta$ with the fluctuation operator
\begin{equation}
\mathcal H_B=-\frac{3\mathcal M}{8\pi^2\kappa}\nabla^2+\xi^{-2}\partial_\Psi^2V(0,\Psi_B).
\end{equation}
The fluctuation operator about the false vacuum, $\mathcal H_F$, follows by replacing $\Psi_B\to\Psi_2$ and all its eigenvalues are positive. In contrast, $\mathcal H_B$ has two zero modes, the translations of the bounce center, and one negative mode, the change of the bubble radius, $d^2B/dR^2|_{R_B}=-2\pi\epsilon<0$. We denote by $\det_+\mathcal H_B$ the product of its remaining, positive eigenvalues.

The two zero modes are traded for an integral over the bubble position, giving the factor $\frac{4\pi\kappa B}{3\mathcal M}\frac{\mathcal A}{\xi^2}$, with $\mathcal A$ the area of the system. The negative mode, with normalization $\sqrt{B/(2\pi\cdot\frac{3\mathcal M}{16\pi^2\kappa})}=\sqrt{8\pi\kappa B/(3\mathcal M)}$, gives the radial integral above and the factor $i$ signaling the instability. The positive modes enter through the ratio $\det\mathcal H_F/\det_+\mathcal H_B$. Measuring eigenvalues in units of $\xi^{-2}$, $\mathcal H_F$ has three more positive eigenvalues than $\det_+\mathcal H_B$, each of order $\partial_\Psi^2V(0,\Psi_2)\simeq2b=\mathcal M^3/(4\pi^2\kappa)$ for $\mathcal Ye^{\Psi_2}\gg1$. Hence
\begin{equation}
\Big[\frac{\det\mathcal H_F}{\det_+\mathcal H_B}\Big]^{1/2} \simeq \Big(\frac{\mathcal M^3}{4\pi^2\kappa}\Big)^{3/2}\mathcal C_b,
\end{equation}
with $\mathcal C_b$ an $O(1)$ numerical constant from the ratio of the remaining spectra.
\subsection{Assembling the decay rate}
Combining the zero-mode factor, the positive-mode determinant ratio, the negative-mode integral and the contribution of the superfluid inside the bubble, the single-bounce amplitude relative to the no-bounce background is $\mathcal Z_1=i\mathcal A(\mathcal X/T)e^{-B}$, with
\begin{align}
\mathcal X={}&\frac{T}{\xi^2}\,\frac{4\pi\kappa B}{3\mathcal M}\,
\sqrt{\frac{8\pi\kappa B}{3\mathcal M}\,\frac{1}{\epsilon\xi^2}}\,
\Big(\frac{\mathcal M^3}{4\pi^2\kappa}\Big)^{3/2}\mathcal C_b\notag\\
&\times e^{Z_D'(0)/2}(4\pi^2K)^{1/12}\Big(\frac{R_B}{\xi}\Big)^{1/6}.
\end{align}
Summing the dilute, non-interacting bounce gas gives $Z_{\rm full}/Z_0=\exp[i\mathcal A(\mathcal X/T)e^{-B}]$, i.e.\ an imaginary part of the free energy per unit area $\propto(\mathcal X/T)e^{-B}$. The quantity $(\mathcal X/T)e^{-B}$ has the dimension of inverse area: it is the density of critical bubbles, fixed by the static action. The decay rate per unit area follows by multiplying it with the growth rate of the unstable radial mode, which is set by the dynamics and not by the static action,  we estimate it as $T$. Using Eqs.~\eqref{eq: bounce action} and \eqref{eq: bounce radius} and collecting powers, up to $O(1)$ numerical prefactors,
\begin{align}
\Gamma\sim{}&\frac{T}{\xi^2}\,K^{1/12}\,\frac{\mathcal M^{28/3}}{\kappa^{19/6}}\,\frac{\ln^{19/3}(1/\mathcal Y)}{(\epsilon\xi^2)^{13/6}}\notag\\
&\times\exp\Big[-\frac{3\mathcal M^4}{32\pi^3\kappa^2}\,\frac{\ln^4(1/\mathcal Y)}{\epsilon\xi^2}\Big].
\end{align}
Inserting the bare values of $\mathcal M$ and $\kappa$, and using $2b\ln(1/\mathcal Y)\simeq\tfrac12$ in the form
\begin{equation}
\ln\frac1{\mathcal Y}=\pi\frac{E_0}{T}\Big(\frac{\xi_F}{\xi}\Big)^2,
\end{equation}
gives, up to $O(1)$ numerical prefactors,
\begin{align}
\Gamma \sim{}&
\frac{T}{\xi^2}\,K^{1/12}\Big(\frac{E_0}{T}\Big)^{19/6}\Big(\frac{\xi_F}{\xi}\Big)^{13/2}\frac{1}{(\epsilon\xi^2)^{13/6}}\notag\\
&\times\exp\Bigg[-\frac{3\pi^3}{8}\Big(\frac{E_0}{T}\Big)^2\Big(\frac{\xi_F}{\xi}\Big)^6\frac{1}{\epsilon\xi^2}\Bigg].
\label{eq:FullGamma}
\end{align}
This concludes the derivation of Eq.~\eqref{eq: decay rate result} of the main text. 
All results of this appendix are obtained within the gradient expansion of Eq.~\eqref{eq: instanton effective action}; as discussed in Sec.~\ref{sec: instantons} and in App.~\ref{ap: scales}, numerical coefficients of $\sigma_\Psi$ and $B$ are therefore reliable only up to factors of order unity.

\bibliography{references}
\end{document}